\documentclass[aps,prl,twocolumn,preprintnumbers,amsmath,amssymb,superscriptaddress]{revtex4-2}
\usepackage{graphicx}
\usepackage[caption=false]{subfig}
\usepackage{epsfig}
\usepackage{dcolumn}
\usepackage{bm}
\usepackage{color}
\usepackage{float}
\usepackage[colorlinks]{hyperref}
\usepackage{verbatim}
\usepackage{algorithm}
\usepackage{algorithmic}
\usepackage{tikz}
\usetikzlibrary{quantikz}

\hypersetup{citecolor = blue}

\def\be{\begin{equation}}       \def\ee{\end{equation}}
\def\bea{\begin{eqnarray}}      \def\eea{\end{eqnarray}}
\def\ba{\begin{array}}
\def\ea{\end{array}}
\def\bnum{\begin{enumerate} }
\def\enum{\end{enumerate}}

\def\=>{\Rightarrow}
\def\>{\rightarrow}

\def\eye2{Fathbb{I}}

\renewcommand{\>}{\rangle}

\renewcommand{\rm}[1]{\mathrm{#1}}

\usepackage{listings}
\usepackage{color}
\definecolor{lightgray}{gray}{1}

\begin{document}

\title{Universal KPZ scaling in noisy hybrid quantum circuits}
%Universal KPZ scaling in noisy hybrid quantum circuits
%KPZ fluctuation from different length scales in noisy hybrid quantum circuits
%Universal quantum noise induced power law entanglement in monitored systems
\author{Shuo Liu}
\altaffiliation{The two authors contributed equally to this work.} 
\affiliation{Institute for Advanced Study, Tsinghua University, Beijing 100084, China}
\author{Ming-Rui Li}
\altaffiliation{The two authors contributed equally to this work.} 
\affiliation{Institute for Advanced Study, Tsinghua University, Beijing 100084, China}
\author{Shi-Xin Zhang}
\email{shixinzhang@tencent.com}
\affiliation{Tencent Quantum Laboratory, Tencent, Shenzhen, Guangdong 518057, China}
\author{Shao-Kai Jian}
\email{skjian@brandeis.edu}
\affiliation{Department of Physics, Brandeis University, Waltham, Massachusetts 02453, USA}
\author{Hong Yao}
\email{yaohong@tsinghua.edu.cn}
\affiliation{Institute for Advanced Study, Tsinghua University, Beijing 100084, China}

\begin{abstract}
Measurement-induced phase transitions (MIPT) have attracted increasing attention due to the rich phenomenology of entanglement structures and their relation with quantum information processing. Since physical systems are unavoidably coupled to environment, quantum noise needs be considered in analyzing a system with MIPT, which may qualitatively modify or even destroy certain entanglement structure of the system. In this Letter, we investigate the effect of quantum noise modeled by reset quantum channel acting on each site with probability $q$ on MIPT. Based on the numerical results from the Clifford circuits, we show that the quantum noise can qualitatively change the entanglement properties - the entanglement obeys ``area law'' instead of ``volume law'' with projective measurement rate $p<p_{c}$. In the quantum noise induced ``area law'' phase, the entanglement exhibits a novel $q^{-1/3}$ power-law scaling. Using an analytic mapping of the quantum model to a classical statistical model, we further show that the ``area law'' entanglement is the consequence of the noise-driven symmetry-breaking field and the $q^{-1/3}$ scaling can be understood as the result of Kardar-Parisi-Zhang (KPZ) fluctuations of the directed polymer with an effective length scale $L_{\rm{eff}} \sim q^{-1}$ in a random environment. 
\end{abstract}

\date{\today}
\maketitle
{\bf Introduction:} The monitored quantum systems undergoing random unitary evolution interspersed by local measurements can present rich entanglement structures. %cite{QI_MIPT_PRB18, MIPT_Li_PRB19, QI_MIPT_PRX19, PhysRevLett.126.060501, PhysRevX.12.011045, PRXQuantum.2.040319, MIPT_Clifford_PRL20_Qi, NonlocalMIPT_PhysRevX}. 
The random unitary evolution generates entanglement within the system while the monitored measurement projects the quantum state to a lower entangled state rendering the system short-range entangled. The competition between unitary evolution and monitored measurement leads to the measurement-induced phase transitions \cite{QI_MIPT_PRB18, MIPT_Li_PRB19, QI_MIPT_PRX19, PhysRevLett.126.060501, PhysRevX.12.011045, PRXQuantum.2.040319, MIPT_Clifford_PRL20_Qi, NonlocalMIPT_PhysRevX,MIPT_Clifford_PRB19, MIPT_Clifford_PRB19_Szyniszewski, MIPT_Clifford_PRL20_Qi, MIPT_Clifford_PRB20_Bao, MIPT_Clifford_PRB21_Fan, MIPT_Clifford_PRB21_Li, MIPT_Clifford_PRB21_Jian}. Below a critical measurement rate, $p_{c}$, the system exhibits large-scale quantum entanglement and enter the  ``volume law'' entanglement phase. Increasing the measurement rate $p$ above the critical rate, the effect of measurements dominates and the entanglement obeys ``area law''. 
The measurement-induced phase transition has also been investigated in the monitored SYK models \cite{MIPT_SYK_1, MIPT_SYK_2, MIPT_SYK_3} and the monitored systems with long-range interactions \cite{NonlocalMIPT_Quantum, NonlocalMIPT_PRXQuantum, NonlocalMIPT_PhysRevX, NonlocalMIPT_PhysRevB, NonlocalMIPT_Qi, NonlocalMIPT_Vijay, MIPT_SYK_1, NonlocalMIPT_PhysRevLett22, NonlocalMIPT_PhysRevLett22_Bao, NonlocalMIPT_PhysRevResearch22, NonlocalMIPT_PhysRevLett22_Minato, NonlocalMIPT_PhysRevX}.

Real physical systems are unavoidably coupled to an environment and thus evolve into mixed states in which von Neumann entropy fails to quantify the quantum entanglement \cite{entropy_fail_1, entropy_fail_2} while the logarithmic entanglement negativity is still a good measure for the mixed-state bipartite entanglement \cite{Negativity_PhysRevA02, Negativity_PhysRevLett05, Negativity_PhysRevLett12, Negativity_Calabrese, Negativity_PhysRevB19, Negativity_PhysRevLett20, Negativity_PhysRevLett20_Wu, Negativity_PhysRevB20, Negativity_PRXQuantum21,Negativity_Shapourian}. The quantum noises and quantum decoherence, induced by the environment, can suppress the entanglement within the systems and are the major obstacles in quantum information processing \cite{QI1_Kim, QI2_Adam, QI3_Nahum, QI_MIPT_PRB18, QI_MIPT_PRX19, QI_nature19,QI_choi, QI_NP, QI_PRX}.  As known before \cite{BAO2021168618, Noise_bulk}, the bulk quantum noises drive the systems to enter the ``area law'' entanglement phase instead of the ``volume law'' phase with $p<p_{c}$, as a consequence of the symmetry-breaking field in terms of the effective statistical model. Nevertheless, there is a novel power law scaling in terms of the system size for the entanglement within the system in the presence of fixed quantum noises at the spatial boundary, in which the quantum noises are modeled by the dephasing channels \cite{Noise_PhysRevLett22}. A straightforward and vital question is whether there is a unified analytic model to understand the effects of the quantum noises of different types and with different space-time distributions.

Despite the similarity of different quantum channels in the large $d$ limit of the classical statistical model, the effect of other quantum channels are remained to be investigated as the quantum systems with qubits (local Hilbert space $d=2$) are the most relevant for quantum information and quantum computation. Besides the dephasing channel, reset can also model the uncontrolled quantum noise in which the $l$-th qudit is reset to $\vert 0 \rangle$ state by reset quantum channel $R_{l}$ and loses correlations with the rest qudits. In addition, the reset channel is easy to implement on the current generation of quantum hardware which is of great experimental relevance as a controlled noise source \cite{reset_exp1, reset_exp2}.

In this Letter, we investigate the entanglement behaviors of the monitored systems in the presence of quantum noise modeled by reset quantum channels. We focus on the case when the probability of measurement $p$ is below the critical probability $p_{c}$, i.e. the system sustains large-scale entanglement in the absence of quantum noise (see \cite{SM} for the results with $p>p_{c}$). To quantify the entanglement within the mixed state, we utilize the logarithmic entanglement negativity $E_{N}$ %\cite{Negativity_PhysRevA02, Negativity_PhysRevLett05, Negativity_PhysRevLett12, Negativity_Calabrese, Negativity_PhysRevB19, Negativity_PhysRevLett20, Negativity_PhysRevLett20_Wu, Negativity_PhysRevB20, Negativity_PRXQuantum21,Negativity_Shapourian}
as discussed above and we also compute the mutual information $I_{A:B}$ which is more intuitive and has similar qualitative properties as logarithmic entanglement negativity.

\begin{figure}[t]
\centering
\begin{minipage}[r]{0.2\textwidth}
\centering
\includegraphics[height=4.5cm,width=4.0cm]{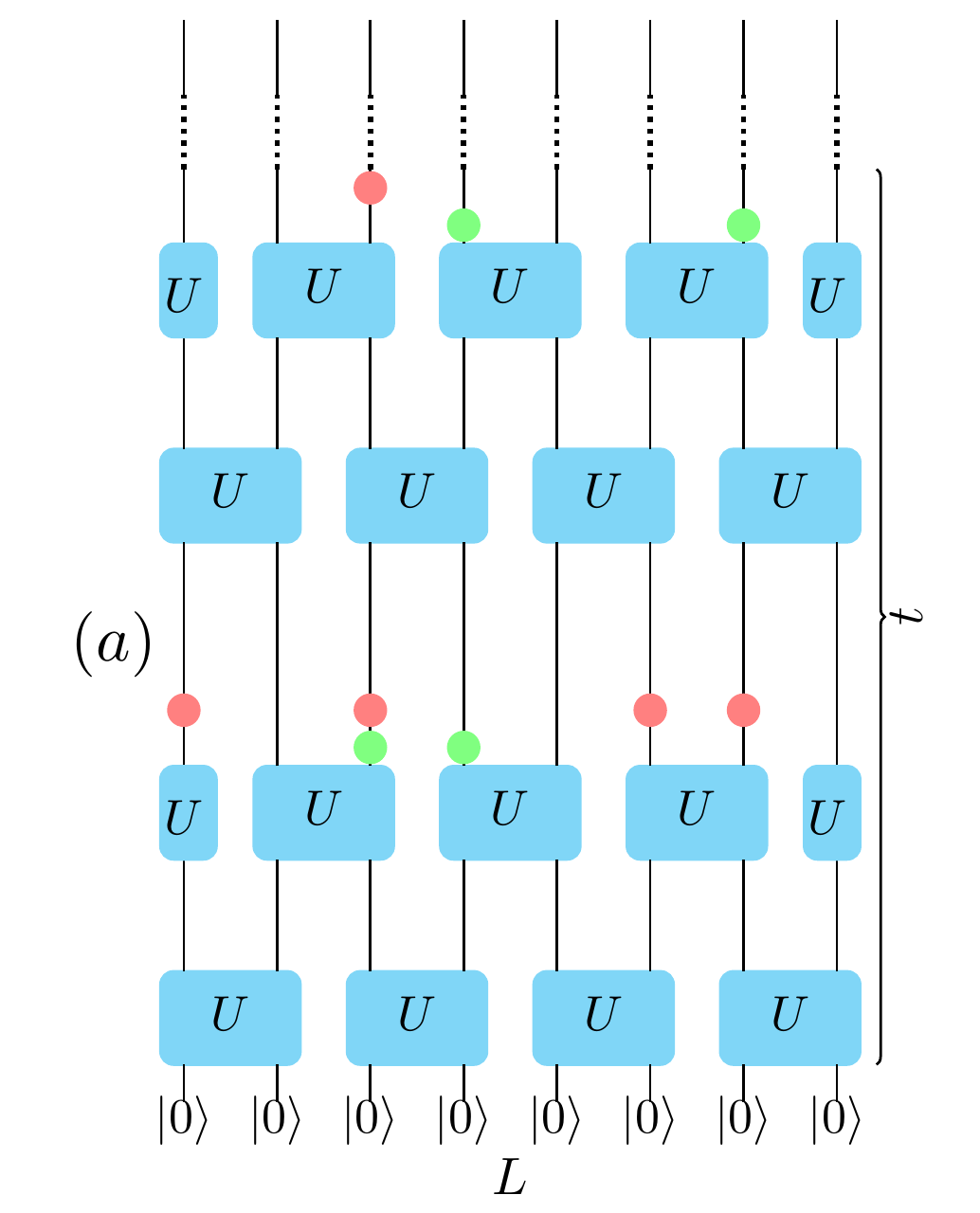}
\end{minipage}%
\begin{minipage}[c]{0.28\textwidth}
\centering
\includegraphics[height=4.5cm,width=4.0cm]{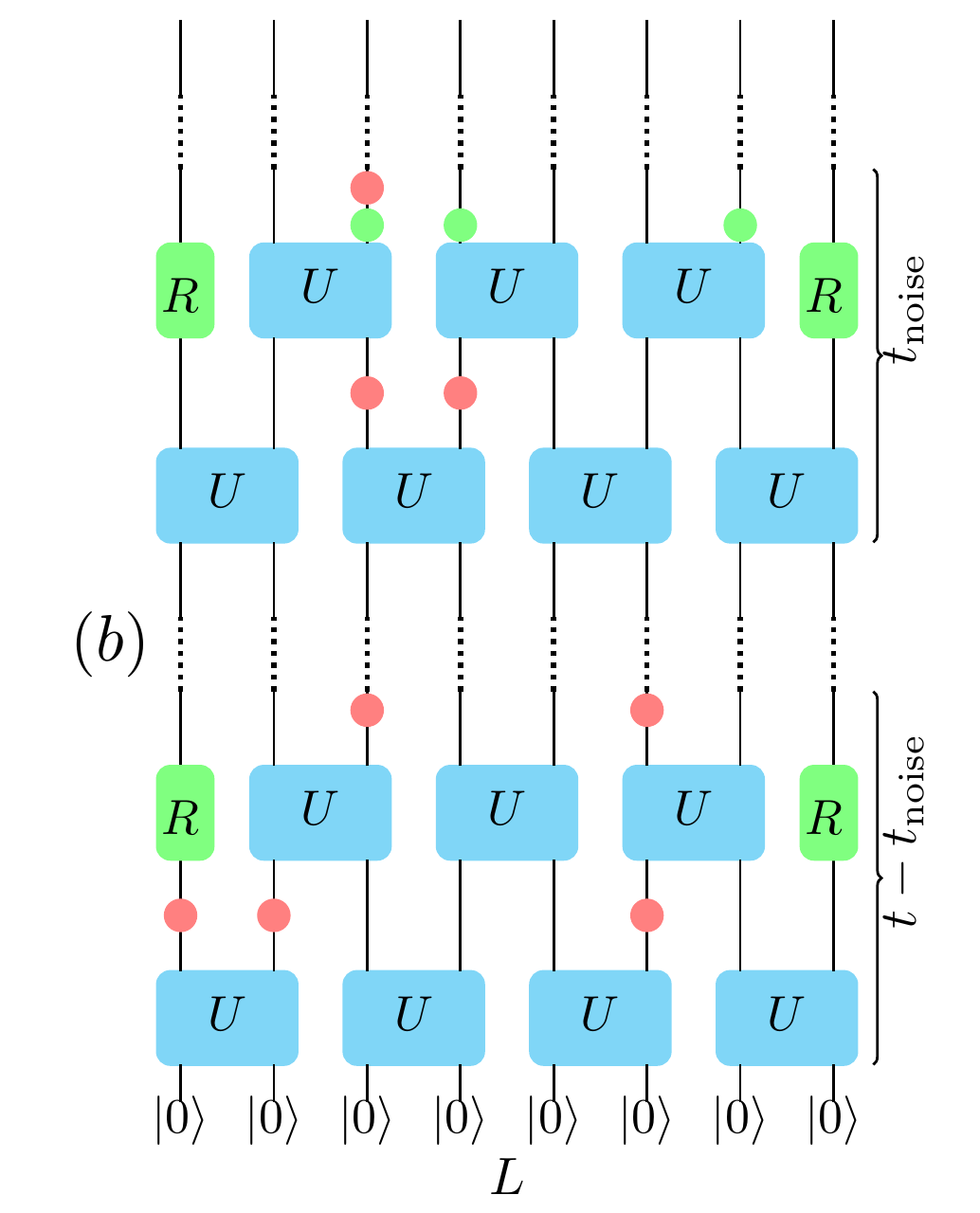}
\end{minipage}
\caption{(a), Circuit diagram in the presence of random bulk resets with $L=8$ qudits and PBC. Qudits are initialized to the product state $\vert 0 \rangle^{L}$ and evolved by applying random uniformly distributed two-qudit Clifford gates (blue blocks). Projective measurements (red dots) occur randomly at a rate $p$ and reset quantum channels (green dots) occur randomly at a rate $q$; (b), Circuit diagram in the presence of fixed resets at the boundary (green blocks) with $L=8$ qudits. On the last $t_{\rm{noise}}$ time steps, the reset quantum channels (green dots) occur randomly at a rate $q$. The rate $q$ and the depth $t_{\text{noise}}$ are adjustable. We set $t=8L$ throughout the work to observe the late-time properties.}
\label{fig:model}
\end{figure}

In the presence of random bulk resets with occurring probability $q$ (see Fig. \ref{fig:model}(a)), the entanglement within the system obeys ``area law'' as predicted by the previous studies \cite{Noise_bulk}. Besides, there is a novel power law scaling for entanglement in terms of the reset probability $q$, $I_{A:B}(q) \sim q^{-1/3}$ and $E_{N}(q) \sim q^{-1/3}$, different from the report in \cite{Noise_bulk}. To deepen the analytic understanding, we map the random circuit evolution to an effective statistical model \cite{Noise_PhysRevLett22}. The mutual information and logarithmic entanglement negativity can be interpreted as the free energy difference of the statistical models with different boundary conditions. And the free energy of the classical statistical model is proportional to the length of the domain wall, i.e., the directed polymer in a random potential. The bulk resets act as the symmetry-breaking field which suppresses the vertical fluctuations of directed polymer and drives the system into the ``area law'' entanglement phase. The resets near the top temporal boundary induce an effective length scale $L_{\text{eff}} \sim q^{-1}$ and the novel power law entanglement in terms of the reset probability $q$ can be understood as the result of Kardar-Paris-Zhang (KPZ) fluctuations of directed polymers \cite{KPZ1, KPZ2, KPZ3} with the effective length scale $L_{\text{eff}}$ instead of the original length scale $L$ \cite{SM}.

\begin{figure}[t]\centering
	\includegraphics[width=0.5\textwidth]{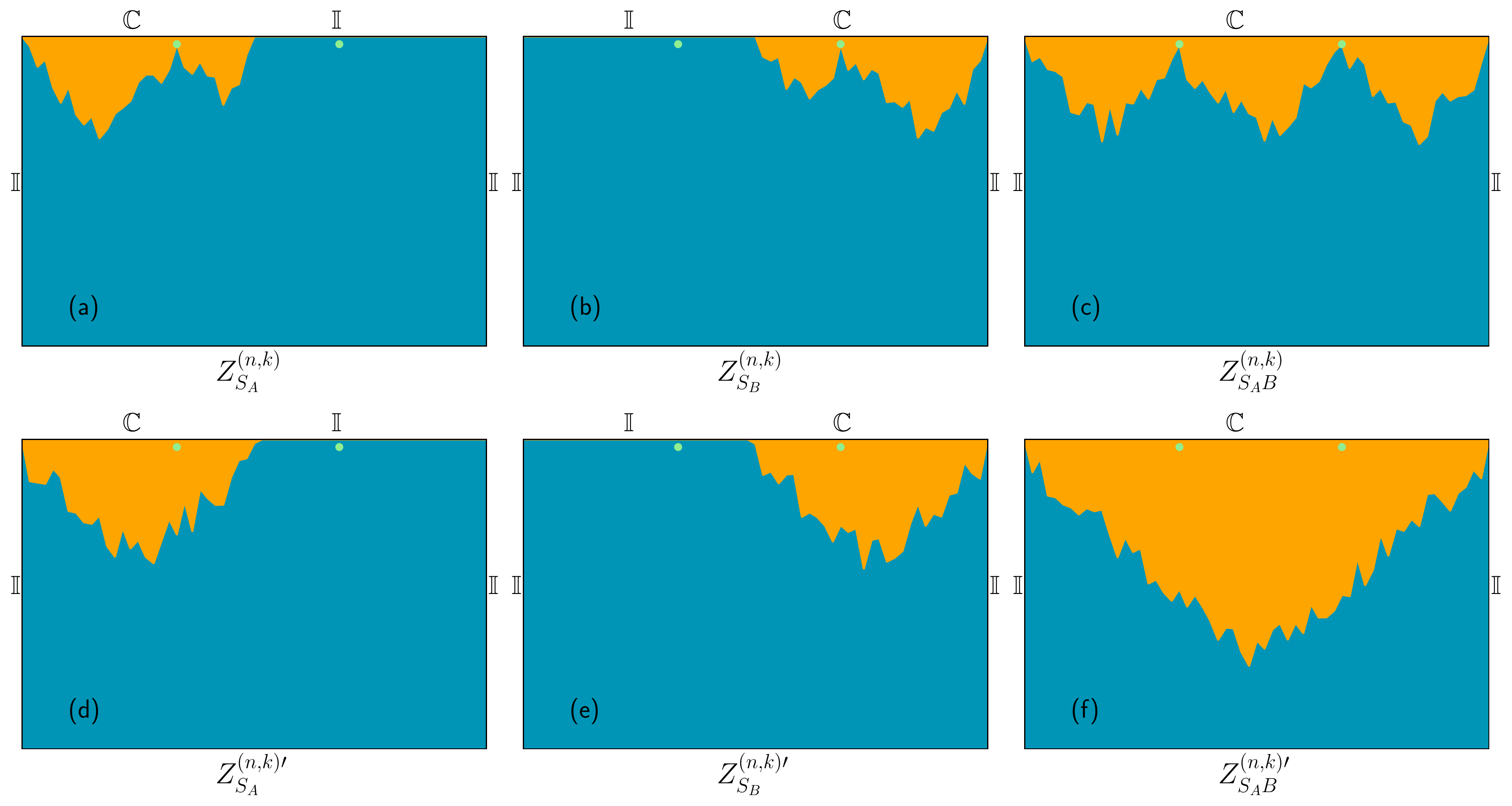}
	\caption{Two possible scenarios for domain configuration in the presence of resets (the resets near the top boundary are represented by green dots): the domain configurations for (a), $Z_{S_{A}}^{(n,k)}$, (b), $Z_{S_{B}}^{(n,k)}$, (c), $Z_{S_{AB}}^{(n,k)}$ in which the vertical fluctuations of the domain wall have been suppressed by resets in the bulk; the domain configurations for (d), $Z_{S_{A}}^{(n,k) \prime}$, (e), $Z_{S_{B}}^{(n,k) \prime}$, (f), $Z_{S_{AB}}^{(n,k) \prime}$ in which the domain wall is not affected by resets in the bulk. For $Z^{(n,k)}$, all the spins are $\mathbb{I}$ and the free energy is zero.  The mutual information is $I_{A:B} =\underset{\underset{k \rightarrow 0}{n \rightarrow 1}}{\lim} \frac{1}{k(n-1)} (F^{(n,k)}_{S_{A}}+F^{(n,k)}_{S_{B}}-F^{(n,k)}_{S_{AB}})$.
	As suggested by the numerical results and the analytic model, the vertical fluctuations are suppressed, and % $Z_{S_{A}}^{(n,k)}$, $Z_{S_{B}}^{(n,k)}$ and $Z_{S_{AB}}^{(n,k)}$
 the upper row are the realistic domain configurations.}
	\label{fig:MI}
\end{figure}

Furthermore, such an analytic model can unify the model with quantum noise at the spatial boundary and in the bulk. To verify this analytic model, we also investigate the entanglement behaviors for the systems with fixed resets at the spatial boundary and random bulk resets on the last $t_{\rm{noise}}$ layers with occurring rate $q$, as shown in Fig. \ref{fig:model}(b). When $t_{\text{noise}}=0$, i.e., zero bulk quantum noise, it is the same as that studied in Ref. \cite{Noise_PhysRevLett22} and exhibits $L^{1/3}$ power law entanglement which is induced by KPZ fluctuations with original length scale $L$ as shown in the lower panel of Fig. \ref{fig:MI}. Via increasing $t_{\rm{noise}}$, i.e., the strength of quantum noise, the entanglement is suppressed and the system enters the ``area law'' phase. In this noise-driven ``area law'' phase, the $q^{-1/3}$ scaling emerges with an effective length scale $L_{\text{eff}}$ as shown in the upper panel of Fig. \ref{fig:MI}. In $t_{\text{noise}} \rightarrow \infty$ limit, it is equivalent to the model shown in Fig. \ref{fig:model}(a) with rescaled reset probability and the $q^{-1/3}$ scaling remains. Based on the analytical understanding, the entanglement behaviors can be unified as $L_{\text{eff}}^{1/3}$, where different space-time distributions of quantum noise induce different $L_{\text{eff}}$.

{\bf Model and observables:} As indicated in Fig. \ref{fig:model}(a), we investigate a one-dimensional system  with $L$ $d$-qudits with initial input state $\vert 0 \rangle ^{L}$. The evolution of the system is determined by a brick-wall random unitary circuit with periodic boundary conditions (PBC) where each gate is independently drawn from the Haar ensemble (or from random two-qubit Clifford ensemble in Clifford simulation). Each single discrete time step consists of four layers. The first two layers are the Haar random unitary two-qudit gates, followed by one layer of reset quantum channels occurring at a rate $q$ on each site $l$ and one layer of projective measurements occurring at a rate $p$ on each site $l^{\prime}$. The reset quantum channel $R_{l}$ on $l$-th qudit takes the density matrix $\rho$ to the mixed state
\bea
\rho^{\prime} = R_{l}[\rho] = \sum_{a=0}^{d-1}E_{l}^{a} \rho E_{l}^{a\dagger},
\eea
where the Kraus operator $E_{l}^{a\dagger}=\vert a \rangle_{l} \langle 0 \vert$.
The projective measurement on $l^{\prime}$-th qubit take the density matrix $\rho$ to $P_{l^{\prime}}^{a} \rho P_{l^{\prime}}^{a\dagger}/\vert\vert P_{l^{\prime}}^{a} \rho P_{l^{\prime}}^{a\dagger}\vert\vert$ with probability $p_a = \vert\vert P_{l^{\prime}}^{a} \rho P_{l^{\prime}}^{a\dagger}\vert\vert$ for $a=0,1,...,d-1$, where $P^{a \dagger}_{l^{\prime}} = P_{l^{\prime}}^{a} = \vert a \rangle_{l^{\prime}} \langle a \vert$.

The quantum entanglement within the system at late times ($t=8L$) is quantified by the logarithmic entanglement negativity
\bea
E_{N} = \text{log} \vert \vert \rho^{T_B} \vert \vert_1,
\eea
where $\rho^{T_B}$ is the partial transpose of $\rho$ in subsystem $B$ and $\vert \vert \cdot \vert \vert_1$ is the trace norm.  $E_{N}$ is a measure of mixed-state bipartite entanglement \cite{Negativity_PhysRevA02, Negativity_PhysRevLett05, Negativity_PhysRevLett12, Negativity_Calabrese, Negativity_PhysRevB19, Negativity_PhysRevLett20, Negativity_PhysRevLett20_Wu, Negativity_PhysRevB20, Negativity_PRXQuantum21,Negativity_Shapourian} where bipartite entanglement entropy fails \cite{entropy_fail_1, entropy_fail_2}. The mutual information obeys qualitatively similar scaling to $E_{N}$ and is more intuitive as shown below. The mutual information between subsystems $A$ and $B$ is given by
\bea
I_{A:B}=S_{A}+S_{B}-S_{AB},
\eea
where $S_{\alpha}$ is the von Neumann entropy ($\alpha=A,B,AB$). We set subsystem $A=[0,L/2]$ and $B=[L/2,L]$ throughout the work.

{\bf Numerical Results with bulk resets:} To avoid the severe finite-size effects, we employ random Clifford unitary gates acting on $d=2$ qubits which can be simulated by classical computers efficiently based on the stabilizer formalism. The Clifford gates form a unitary 3-designs \cite{Unitary3design, bergSimpleMethodSampling2021} and thus are expected to give the same qualitative entanglement behaviors as the Haar random circuit. And the entanglement in the thermodynamic limit $L \rightarrow \infty$ can be extrapolated by assuming $S(L,q) = c(q) L^{-1} + S(\infty, q)$ ($S$ is $I_{A:B}$ or $E_{N}$).

\begin{figure}[t]\centering
	\includegraphics[width=0.5\textwidth]{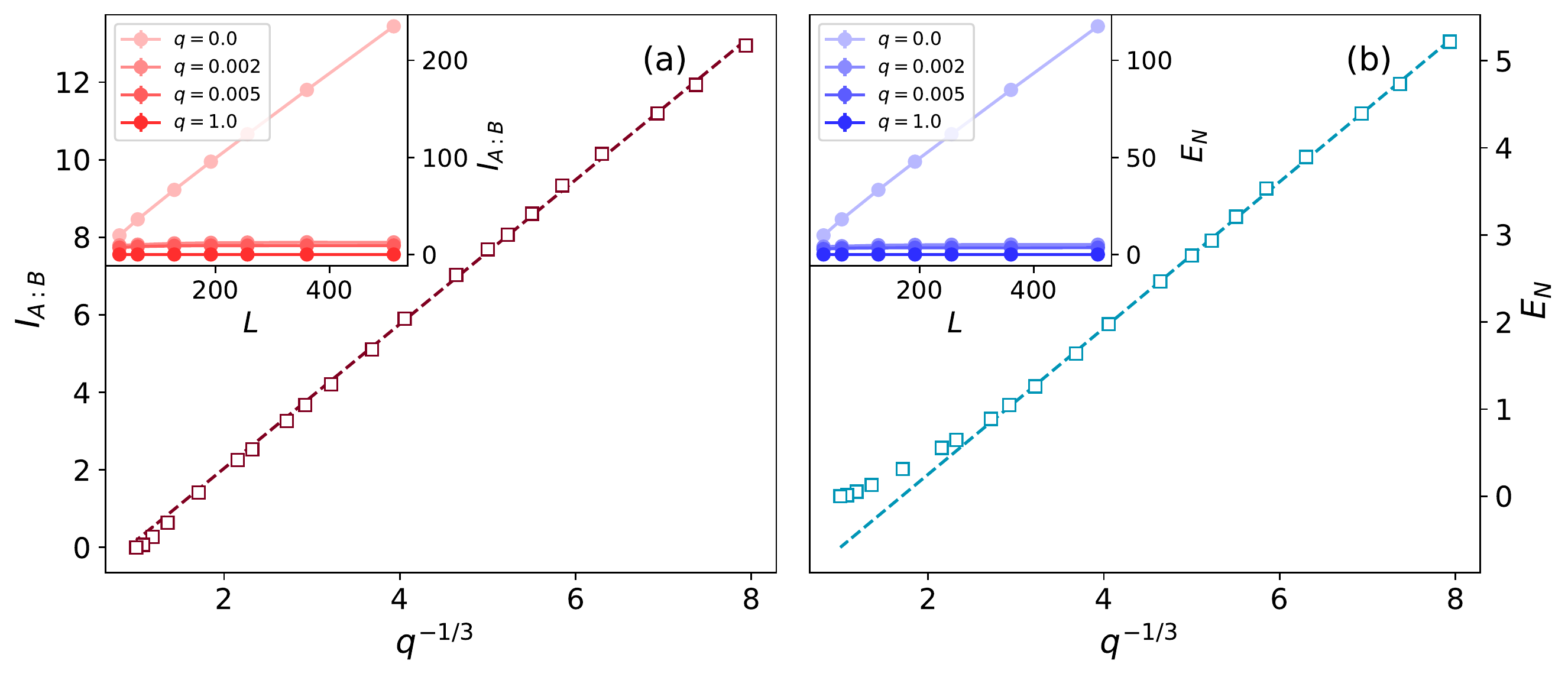}
	\caption{(a) The mutual information $I_{A:B}(q)$ and (b) logarithmic entanglement negativity $E_{N}(q)$ with PBC. The measurement rate is $p=0.1 < p_{c}$. There is a novel $q^{-1/3}$ scaling for the entanglement within the system. The inset is the relationship between $I_{A:B}$ or $E_{N}$ and the system size $L$ with different reset rates. In the absence of reset quantum channels, the entanglement obeys ``volume law''; in the presence of reset quantum channels, the entanglement obeys ``area law''. As $q$ is close to $1$, the entanglement deviates from the predicted value from $q^{-1/3}$ scaling, which can be explained as the breakdown of KPZ field theory due to the small effective length scale.}
	\label{fig:pm0.1PBC}
\end{figure}

For the monitored systems shown in Fig. \ref{fig:model}(a) without resets, i.e., $q=0$, the critical measurement rate is $0.30<p_{c}<0.31$ \cite{Noise_bulk}. Below the critical measurement rate $p_{c}$, the entanglement within the system obeys ``volume law'', i.e., $I_{A:B}(L) \sim L$ and $E_{N}(L) \sim L$, as shown in the insets of Fig. \ref{fig:pm0.1PBC}. With increasing the measurement rate $p$ above $p_{c}$, the system enters the ``area law'' entanglement phase, i.e., $I_{A:B}(L) \sim L^{0}$ and $E_{N}(L) \sim L^{0} $. When the quantum noises induced by the environment and modeled by reset channels are added into the circuit with probability $q$, we focus on the case with $p<p_{c}$ and set $p=0.1$ which is deep in the original ``volume law'' phase (see \cite{SM} for the results with $p>p_{c}$). There is a novel power law scaling in terms of $q$: $I_{A:B}(q) \sim q^{-1/3}$ and $E_{N}(q) \sim q^{-1/3}$, besides the expected ``area law'' entanglement phase in terms of the system size, as indicated in Fig. \ref{fig:pm0.1PBC}. The emergent $q^{-1/3}$ scaling can be understood as the consequence of the KPZ fluctuations with an effective length scale $L_{\text{eff}} \sim q^{-1}$ as discussed below \cite{SM}.

When $q=1$, the steady state is exactly the product state $\vert 0 \rangle^L$ and $I_{A:B}=E_{N}=0$ as shown in Fig. \ref{fig:pm0.1PBC}. As $q$ is close to $1$ and thus the effective length scale is of the same order as the discrete lattice constant, the KPZ field theory description breaks down and the entanglement deviates from the predicted value based on the $q^{-1/3}$ scaling. The choice of boundary condition doesn't change the entanglement behaviors qualitatively (see the numerical results with open boundary condition (OBC) in SM). The entanglement with PBC is about twice as large as that with OBC in the noise-driven ``area law'' phase.

{\bf Effective statistical model:} The numerical results in this Letter can be well explained by a mapping to an effective statistical model. We build upon the previous works \cite{Noise_PhysRevLett22} for a unified analytical picture to understand the noise effects in the monitored random circuit.

To compute the entanglement from the effective statistical model, we consider the $n$th R$\rm{\acute{e}}$nyi entropy $S^{(n)}_{\alpha}$ firstly. For fixed sets of measurement locations $X$ and reset locations $Y$ in the circuit, averaged over Haar unitary $U = \{ U_{ij,t}\}$ and measurement results $m$, $S^{(n)}_{\alpha}$ is
\bea
&&\overline{S^{(n)}_{\alpha}(X,Y)} \\ \nonumber
 && = \mathbb{E}_{U} \sum_{m} p_{m, X, Y} \frac{1}{1-n} \rm{log} \left\{\frac{\rm{tr}(\rho_{\alpha, m, X, Y}^{n})}{(\rm{tr} \rho_{m, X, Y})^{n}} \right\},
\eea
where $\alpha$ denotes the subsystem ($\alpha=A,B,AB$), $\rho_{m, X, Y}$ is the unnormalized density matrix given the measurement trajectory $m$, with probability $p_{m, X, Y}=\rm{tr}\rho_{m, X, Y}$. 

% is the probability for achieving the measurement outcomes $m$ conditioned on the locations of measurements $X$, the locations of resets $Y$ and the unitary realization $U$.
% The entanglement entropy is given by $S_{\alpha} = \lim_{n \rightarrow 1} S_{\alpha}^{(n)}$ and mutual information is defined as $I_{A:B} = \lim_{n \rightarrow 1} (S_{A}^{(n)} + S_{B}^{(n)} - S_{AB}^{(n)})$. The logarithmic entanglement negativity can also be obtained from the $n$th R$\rm{\acute{e}}$nyi negativity similarly which we defer the derivation in the Supplemental Materials.

The average of the log function can be evaluated via the replica trick \cite{replica1, replica2}. To this end, we first perform the average over unitary realizations inside the log function
\bea
S_{\alpha}^{(n,k)} &=& \frac{1}{k(1-n)} \text{log} \left\{ \frac{Z_{S_{\alpha}}^{(n,k)}}{Z^{(n,k)}} \right\}, \\ \nonumber
&=& \frac{1}{k(n-1)} (F^{(n,k)}_{S_{\alpha}} - F^{(n,k)}).
\eea
We map the hybrid circuit with replica trick to an effective statistical model with classical spin freedom that valued over permutation group $S_{nk+1}$ with ferromagnetic spin-spin interaction \cite{SM}. $Z$ are the partition functions of the statistical models with different top boundary conditions: $Z^{(n,k)}$ contains identity permutations $\mathbb{I}$ along the entire top boundary while $Z^{(n,k)}_{S_{\alpha}}$ contains cyclic permutation $\mathbb{C}$ at the top region $\alpha$ and identity permutation $\mathbb{I}$ at the top complementary region (see Fig. \ref{fig:MI}).
% And $F$ are free energies of the statistical models.
% Due to the ferromagnetic spin-spin interaction, the corresponding free energy $F$ is proportional to the length of domain wall.
The mutual information is the difference of free energies $F$ of statistical models with specific boundary conditions
\bea
I_{A:B} &=& \underset{\underset{k \rightarrow 0}{n \rightarrow 1}}{\lim} (S^{(n,k)}_{A} + S^{(n,k)}_{B} - S^{(n,k)}_{AB}), \\ \nonumber
&=& \underset{\underset{k \rightarrow 0}{n \rightarrow 1}}{\lim} \frac{1}{k(n-1)} (F^{(n,k)}_{S_{A}}+F^{(n,k)}_{S_{B}}-F^{(n,k)}_{S_{AB}}).
\eea
The logarithmic entanglement negativity can also be obtained from the replica negativity similarly which we defer the derivation in the Supplemental Materials.

% Utilizing the replica trick \cite{replica1, replica2} to write $\overline{S_{\alpha}^{(n)}} = \text{lim}_{k\rightarrow 0} S_{\alpha}^{(n, k)}$, and $S_{\alpha}^{(n,k)}$ can be interpreted as the free energies difference:
% \bea
% S_{\alpha}^{(n,k)} = \frac{1}{k(1-n)} \text{log} \left\{ \frac{Z_{S_{\alpha}}^{(n,k)}}{Z^{(n,k)}} \right\},
% \eea
% where
% \bea
% Z_{S_{\alpha}}^{(n,k)} &=& \text{Tr} \left\{  (\Sigma_{S_{\alpha}}^{\otimes k} \otimes \mathbb{I}) \mathbb{E}_{U} \sum_{m} \rho_{m}^{\otimes r} \right\}, \\
% Z^{(n,k)} &=& \text{Tr} \left\{  \mathbb{I}^{\otimes r} \mathbb{E}_{U} \sum_{m} \rho_{m}^{\otimes r} \right\},
% \eea
% with $r=nk+1$. $\Sigma$ denotes the corresponding permutation and the partition functions differ only in the top boundary conditions at the final time slice. As shown in Fig. \ref{fig:MI}, $Z^{(n,k)}$ contains identity permutations $\mathbb{I}$ along the entire top boundary while $Z^{(n,k)}_{S_{\alpha}}$ contains cyclic permutation $\mathbb{C}$ at the top region $\alpha$ and identity permutation $\mathbb{I}$ at the top complementary region. \textcolor{blue}{[less technical replica discussion and more description of statistical model? introduce the free energy $F$]}

In the large $d \rightarrow \infty$ limit, the free energy of the effective statistical model is determined by the most probable classical spin configuration and proportional to the domain wall length due to the ferromagnetic spin-spin interaction. In the absence of measurements and resets, the domain wall is unique for the statistical model with specific boundary conditions due to the unitary constrain and the length is proportional to $L$. The measurements act as the pointwise attractive potential and the randomness of measurement locations can be regarded as the quenched disorders. In the coarse-grained picture, the effective statistical model is equivalent to the model of directed polymers in a random Gaussian potential described by the KPZ field theory \cite{KPZ1, KPZ2, KPZ3,Noise_PhysRevLett22}. Thus the directed polymer, i.e., the domain wall, in the random monitored measurement background, fluctuates slightly away from the unique trajectory and the domain wall length is thus $s_0 L + s_1 L^{1/3}$. As predicted by the KPZ field theory, the length scale of the vertical fluctuations of the directed polymer is $O(L^{2/3})$.

The reset quantum channels in the bulk act as a symmetry-breaking field after mapping to the statistical model and the free energy is minimized when the classical spin permutation freedom is pinned to identity $\mathbb{I}$. Due to the non-identity spin permutation freedom induced by the top boundary $\alpha$, the free energy cost is proportional to the number of the resets contained between the domain wall and the top boundary $\alpha$. To avoid this cost, the length scale of vertical fluctuations of the domain wall can be suppressed to exclude more resets. Equivalently, the reset quantum channels in the bulk can be interpreted as attractive potential from the top boundary and can induce the pinning phase transition, where the $O(L^{2/3})$ KPZ vertical fluctuations with length scale $L$ vanishes and the system enters the pinned phases, i.e., ``area law'' entanglement phase  \cite{Pining, Pining_LYD}. Besides, the resets near the top boundary can further induce an effective length scale $L_{\text{eff}} \sim q^{-1}$ and open a possible way for the directed polymer to fluctuate vertically with the emergent and smaller length scale $L_{\text{eff}}$ as indicated in Fig. \ref{fig:MI}. And the domain wall length is now $ s_{0} L_{\text{eff}} + s_{1}L_{\text{eff}}^{1/3}$ due to the KPZ fluctuation \cite{KPZ1, KPZ2, KPZ3,Noise_PhysRevLett22}.

The two possible scenarios discussed above are summarized in Fig. \ref{fig:MI}.
% in which the effective length scale is $L_{\text{eff}}=L/3$.
% The mutual information is the free energy difference of statistical models with specific boundary conditions \textcolor{blue}{[use a different symbol for Renyi mutual information?]}
% \bea
% I_{A:B} = \frac{1}{k(n-1)} (F^{(n,k)}_{S_{A}}+F^{(n,k)}_{S_{B}}-F^{(n,k)}_{S_{AB}}).
% \eea
For $Z^{(n,k)}_{S_{AB}}$, if the two endpoints of the directed polymer are in the same region ($A$ or $B$), the free energy contribution is canceled by the directed polymer in $Z^{(n,k)}_{S_{A}}$ or $Z^{(n,k)}_{S_{B}}$ and the contribution to the mutual information is zero; if the two endpoints are in the region $A$ and $B$ respectively as the middle directed polymer shown in Fig. \ref{fig:MI}(c), the contribution to the mutual information is proportional to $L_{\text{eff}}^{1/3} \sim q^{-1/3}$ \cite{Noise_PhysRevLett22, SM}. Therefore, the novel power law scaling in terms of reset probability $q$ shown in Fig. \ref{fig:pm0.1PBC} can be understood as the consequence of the KPZ fluctuation with an emergent effective length scale $L_{\text{eff}} \sim q^{-1}$. And it is straightforward that the entanglement with PBC is twice as OBC case because the directed polymer can cross the side boundary as well as the middle point, which has a nonzero contribution to the mutual information. In the absence of monitored measurements, i.e., the random attractive potential, the $q^{-1/3}$ scaling disappears as shown in the Supplemental Materials.%Fig. \ref{fig:pm0.0OBC}. % the

\begin{figure}[t]\centering
	\includegraphics[width=0.5\textwidth]{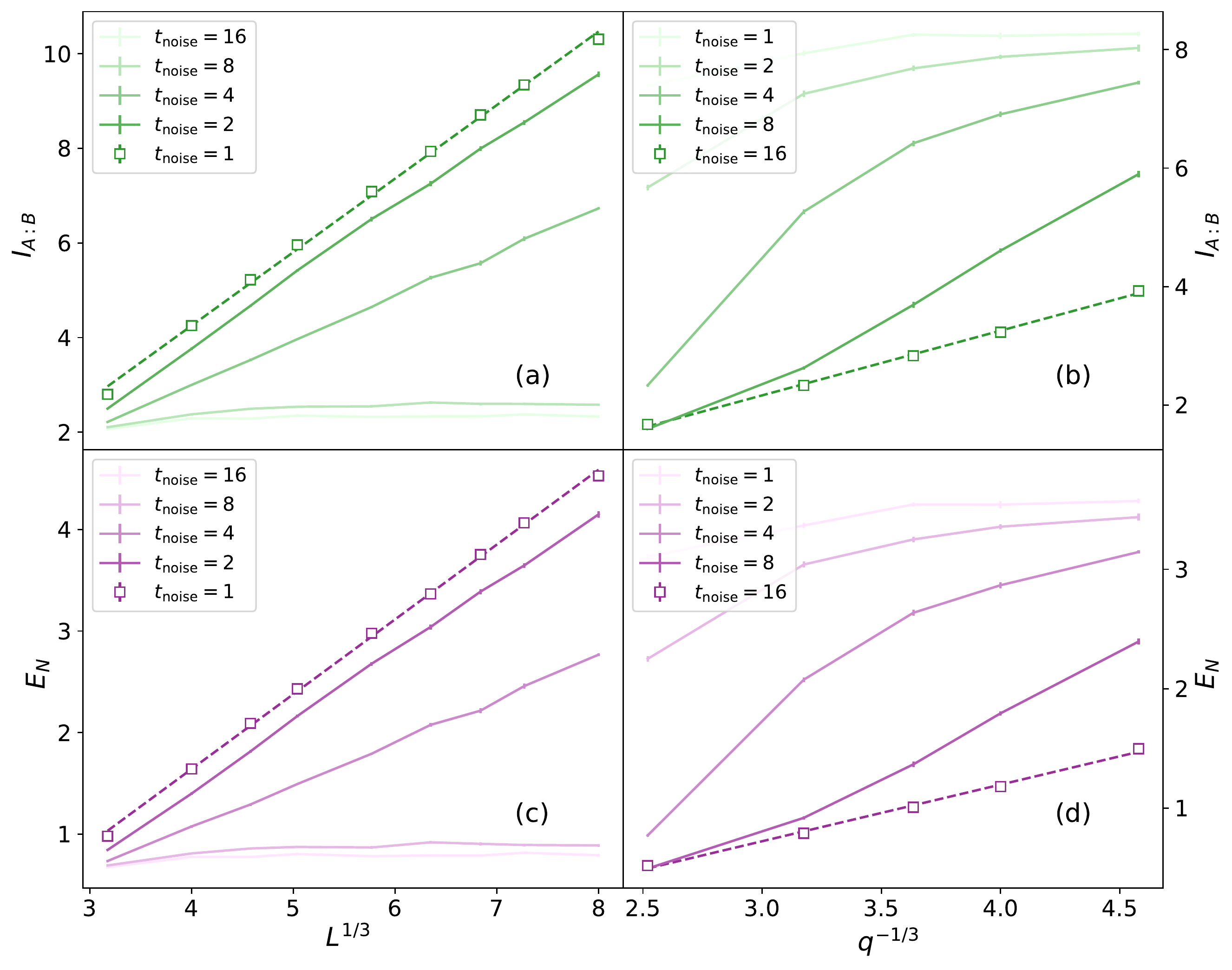}
	\caption{(a), The mutual information $I_{A:B}$ and (c), Entanglement negativity $E_{N}$ with fixed $q=1/32$ and different $t_{\rm{noise}}$. When $t_{\rm{noise}}$ is small, the quantum noise is not enough to suppress the $L^{1/3}$ entanglement by inducing the pinning phase transition, and the $L^{1/3}$ scaling still exists; when $t_{\rm{noise}}$ is large, the system enters the ``area law'' entanglement phase. (b), The mutual information $I_{A:B}$ and (d), Entanglement negativity $E_{N}$ with fixed system size $L=256$ and different $t_{\rm{noise}}$. When $t_{\rm{noise}}$ is large, the ``volume law'' entanglement vanishes and the $q^{-1/3}$ scaling appears.}
	\label{fig:diff4}
\end{figure}

To further verify the statistic model mapping and the analytic picture, we consider another model with both monitored measurements and reset quantum channels. In this model, we have resets on the spatial boundary of the hybrid circuit and also random bulk resets only on the last $t_{\text{noise}}$ layers as shown in Fig. \ref{fig:model}(b). This model is the same as \cite{Noise_PhysRevLett22} with $q=0$ limit or $t_{\text{noise}}=0$ limit (with dephasing channel replaced by reset channel). In the $t_{\text{noise}} \rightarrow \infty$ limit, this model is the same as that shown in Fig. \ref{fig:model}(a) with rescaled $p$ and $q$. Based on the analytic picture discussed above, the resets occurring at a small rate $q$ are not enough to suppress the $O(L^{2/3})$ vertical fluctuations and the systems exhibit power law scaling $(L^{1/3})$ entanglement when $t_{\text{noise}}$ is small. With increasing the $t_{\text{noise}}$, the $O(L^{2/3})$ vertical fluctuation vanishes and the system enters the ``area law'' entanglement phase with a novel power law scaling $q^{-1/3}$ as indicated in Fig. \ref{fig:diff4}. Interpolated by this model, the boundary quantum noise and the bulk quantum noise are unified. In terms of the statistical model, they both play a role in fixing the endpoints of the directed polymer and thus induce length scales $O(L)$ and $O(q^{-1})$, respectively.

Another strategy to detect this entanglement phase transition and verify the analytic picture is by increasing $q$ with fixed $t_{\text{noise}}$, similar to the pinned phase transition setup investigated in \cite{Pining_LYD}. This approach has also been studied and more details can be found in \cite{SM}.

{\bf Conclusions and discussions} The reset quantum channels can drive the systems to enter the ``area law'' entanglement phase as the consequence of the symmetry-breaking in the effective statistical model, similar to other decoherence quantum channels investigated before (eg. dephasing channels and depolarizing channels). More importantly, we identify that there is a novel power-law scaling ($q^{-1/3}$) in the quantum noise-driven ``area law'' phase as the result of KPZ fluctuations with an effective length scale $L_{\text{eff}} \sim q^{-1}$. This new analytic picture, supported by convincing numerical results from models with different space-time distributions of quantum noises, unifies the understanding of boundary and bulk quantum noise in which the difference is the effective length scale induced by the distribution of quantum noise.

Since all decoherence quantum channels break the permutation symmetry in the effective statistical model, we believe that the novel power-law scaling ($q^{-1/3}$) remains in the presence of other quantum noises as a universal behavior in noisy hybrid circuits. Moreover, as indicated by the $q^{-1/3}$ scaling, an interesting future direction is to investigate whether we can identify non-trivial entanglement structure using local probes even in the noise-driven ``area law'' phase.

% \textcolor{blue}{[the following summary is duplicated. replace by discussion of general impact of our work and future direction. 1. Local entanglement structure, entanglement length scale $1/q$. 2. Universal noisy gate.]}
% In terms of the effective statistical model, an effective length scale ($L_{\text{eff}} \sim q^{-1}$) emerges due to the pinning effect of the quantum channel near the top boundary. The novel power law scaling entanglement can be understood as the KPZ fluctuation of the directed polymer with the effective length scale $L_{\text{eff}}$. To verify this analytic picture, we have demonstrated this novel scaling with different boundary conditions (PBC and OBC), with different measurement rates, and with different model setups (see Fig. \ref{fig:model}) based on convincing numerical results. The proposed analytical picture can unify the understanding of boundary and bulk quantum noise in which the difference is the effective length scale induced by the quantum noise locations. With boundary quantum noise, $L_{\text{eff}}$ is exactly the system size and the system exhibits $O(L^{1/3})$ entanglement. With bulk quantum noise, $L_{\text{eff}} \sim q^{-1}$, which is independent of $L$ and the system enters the ``area law'' entanglement phase in which the novel $q^{-1/3}$ scaling appears.

~\newline
\textbf{Acknowledgements:} This work is supported in part by the NSFC under Grant No. 11825404 (SL, MRL, and HY), the MOSTC Grants No. 2018YFA0305604 and No. 2021YFA1400100 (HY) the CAS Strategic Priority Research Program under Grant No. XDB28000000 (HY). The work of SKJ is supported by the Simons Foundation via the It From Qubit Collaboration. 

\bibliographystyle{apsreve}
\bibliography{ref}

\clearpage

\begin{widetext}
	\section*{Supplemental Materials}
	\renewcommand{\theequation}{S\arabic{equation}}
	\setcounter{equation}{0}
	\renewcommand{\thefigure}{S\arabic{figure}}
	\setcounter{figure}{0}

	\subsection{A. Open boundary condition}
	The numerical results in the presence of random reset quantum channels in the bulk %(see Fig. \ref{fig:model}(a))
	with open boundary condition (OBC) are shown in Fig. \ref{fig:pm0.1OBC}. Here, we choose the measurement rate $p=0.1$. There is also a novel power-law scaling ($q^{-1/3}$) which is the same as the PBC case. And the entanglement with PBC is roughly twice as large as that with OBC in the quantum noise-driven ``area law'' entanglement phase. From the effective statistical model picture as discussed below, the domain wall can cross the side boundary as well as the middle point to make the endpoints in the region $A$ and $B$ respectively with PBC, while the directed polymer crosses the side boundary is forbidden with OBC. The numerical results are consistent with the prediction from the analytic model and further demonstrate the validity of our analytic picture.

	\begin{figure}[H]\centering
	\includegraphics[width=0.5\textwidth]{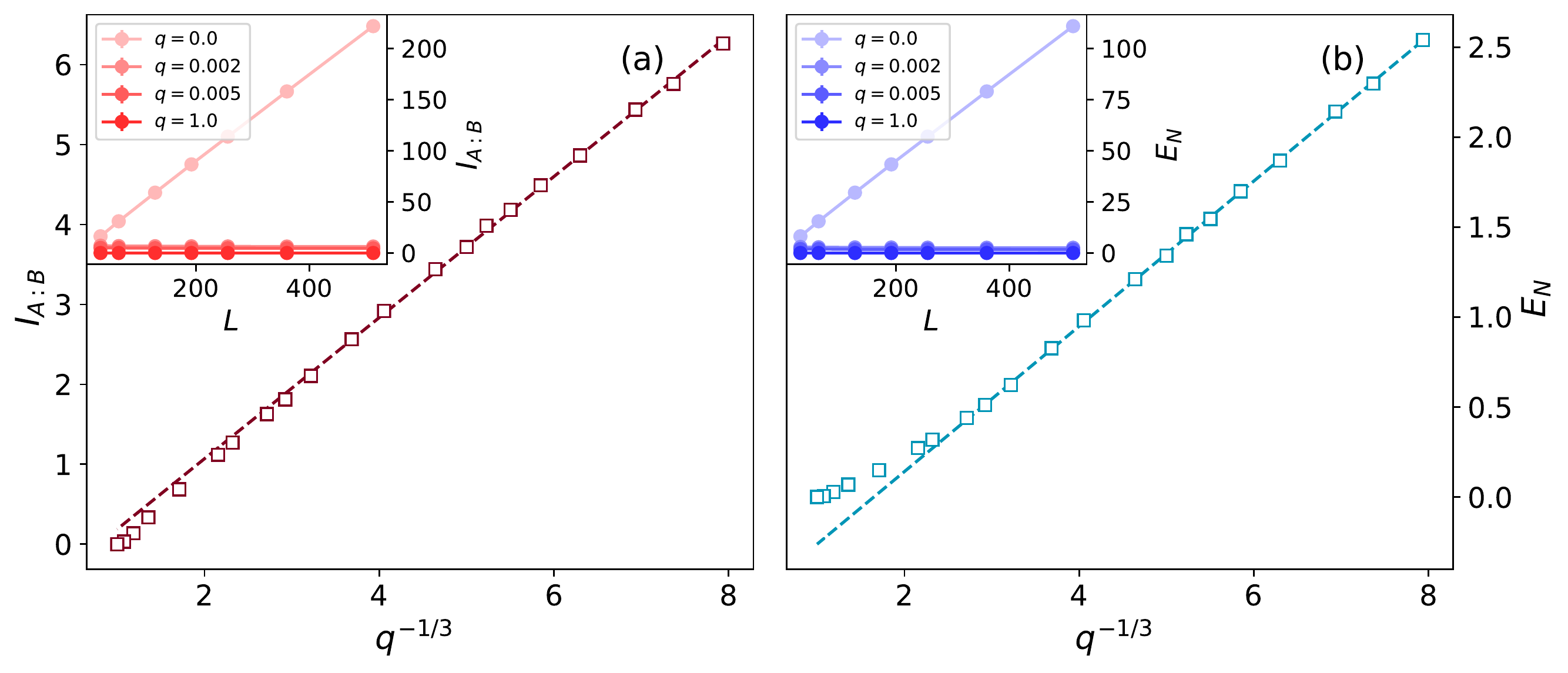}
	\caption{(a), The mutual information $I_{A:B}$ and (b), logarithmic entanglement negativity $E_{N}$ after extrapolating the system size to infinity (OBC). The measurement rate is $p=0.1$. The qualitative behaviors of entanglement with OBC and PBC are essentially the same. When $q$ is finite, entanglement with PBC is roughly twice as large as that with OBC, which is clear in terms of the effective statistical model.}
	\label{fig:pm0.1OBC}
	\end{figure}

	\subsection{B. Large measurement rate: $p>p_{c}$}
	In the main text, we focus on the case with $p<p_{c}$. In this section, we investigate the entanglement behaviors with measurement rate $p$ above the critical rate $p_{c}$ in the presence of reset quantum channels. Here, we set $p=0.6$. The numerical results with reset quantum channels are shown in Fig. \ref{fig:largep}. The entanglement within the system still obeys “area law” when $p>p_{c}$. And there is a plateau with varying $q$ similar to that reported in Ref. \cite{Noise_bulk}: increasing the rate $q$ of quantum noise, there exists a finite region in which the mutual information and logarithmic entanglement negativity don't change and are equal to those in the absence of reset channels.
	\begin{figure}[H]
	\centering
	\begin{minipage}[r]{0.5\textwidth}
	\centering
	\includegraphics[height=4.0cm,width=9cm]{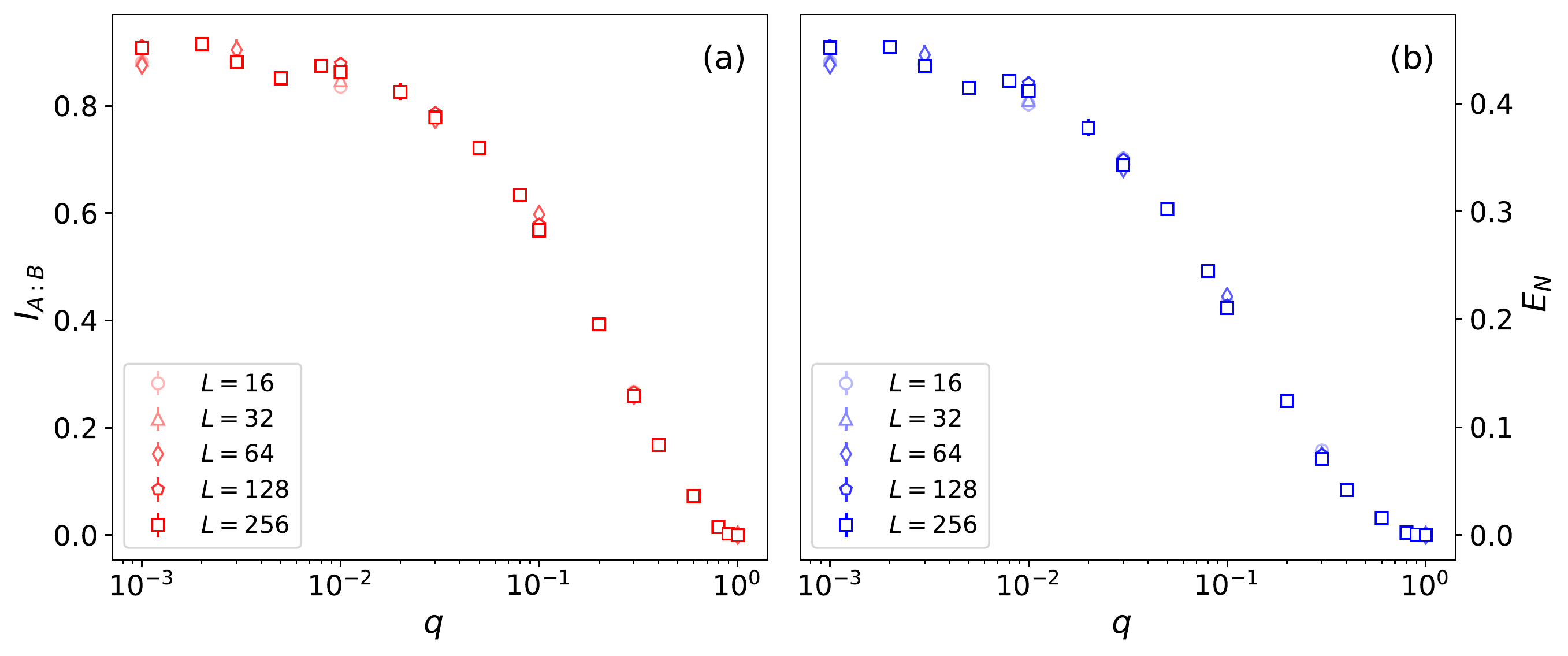}
	\end{minipage}%
	\begin{minipage}[c]{0.5\textwidth}
	\centering
	\includegraphics[height=4.0cm,width=9.5cm]{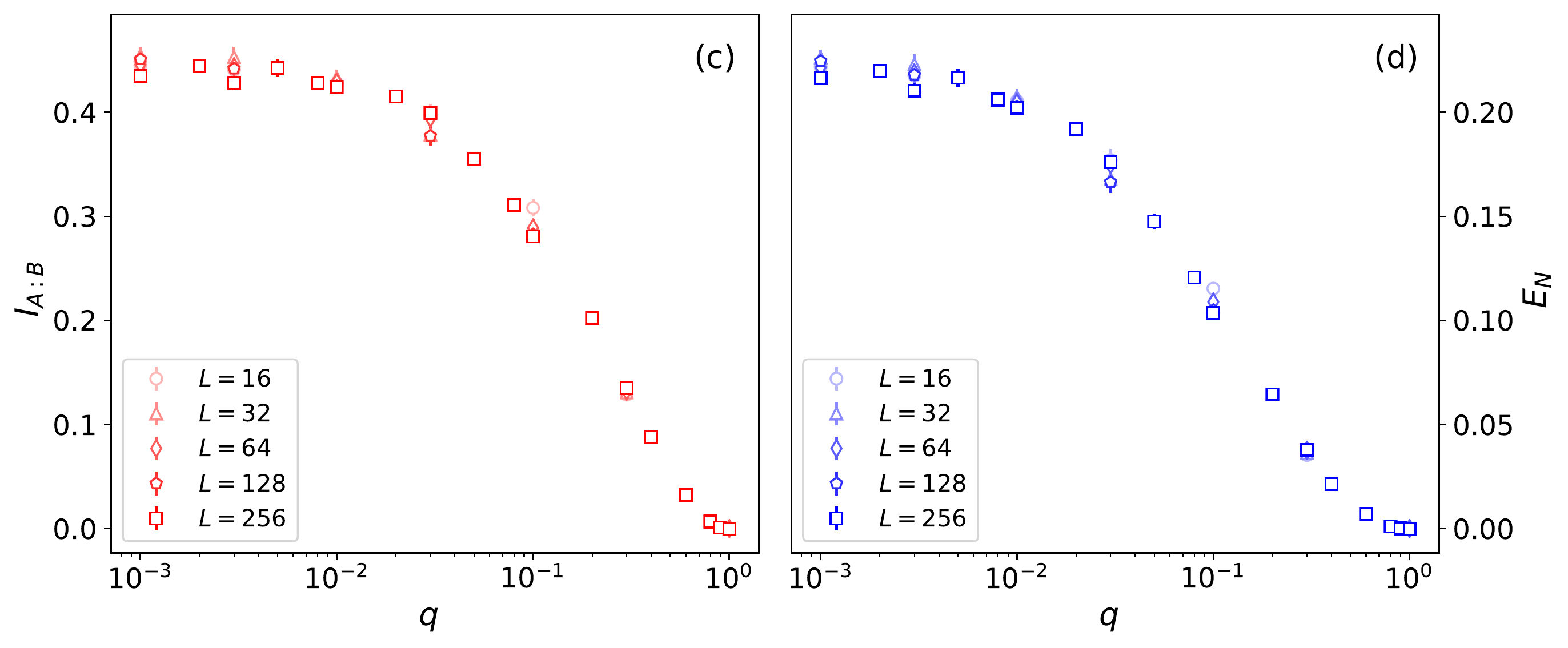}
	\end{minipage}
	\caption{(a), The mutual information $I_{A:B}$ and (b), logarithmic entanglement negativity $E_{N}$ (PBC). (c), The mutual information $I_{A:B}$ and (d), logarithmic entanglement negativity $E_{N}$ (OBC). The measurement rate is $p=0.6$ which is larger than $p_{c}$. The system remains in the ``area law'' entanglement phase in the presence of the reset quantum channels and there is no $q$ dependence when $q$ is small.}
	\label{fig:largep}
	\end{figure}

	\subsection{C. Zero measurement rate: $p=0$}
	In the main text, the $q^{-1/3}$ scaling is explained as the result of KPZ fluctuations of the directed polymers, in which the existence of measurements is necessary. In the absence of measurements, the KPZ field theory fails to describe the effective statistical model and thus the novel $q^{-1/3}$ scaling disappears as shown in Fig. \ref{fig:pm0.0OBC}.
	\begin{figure}[H]\centering
	\includegraphics[width=0.5\textwidth]{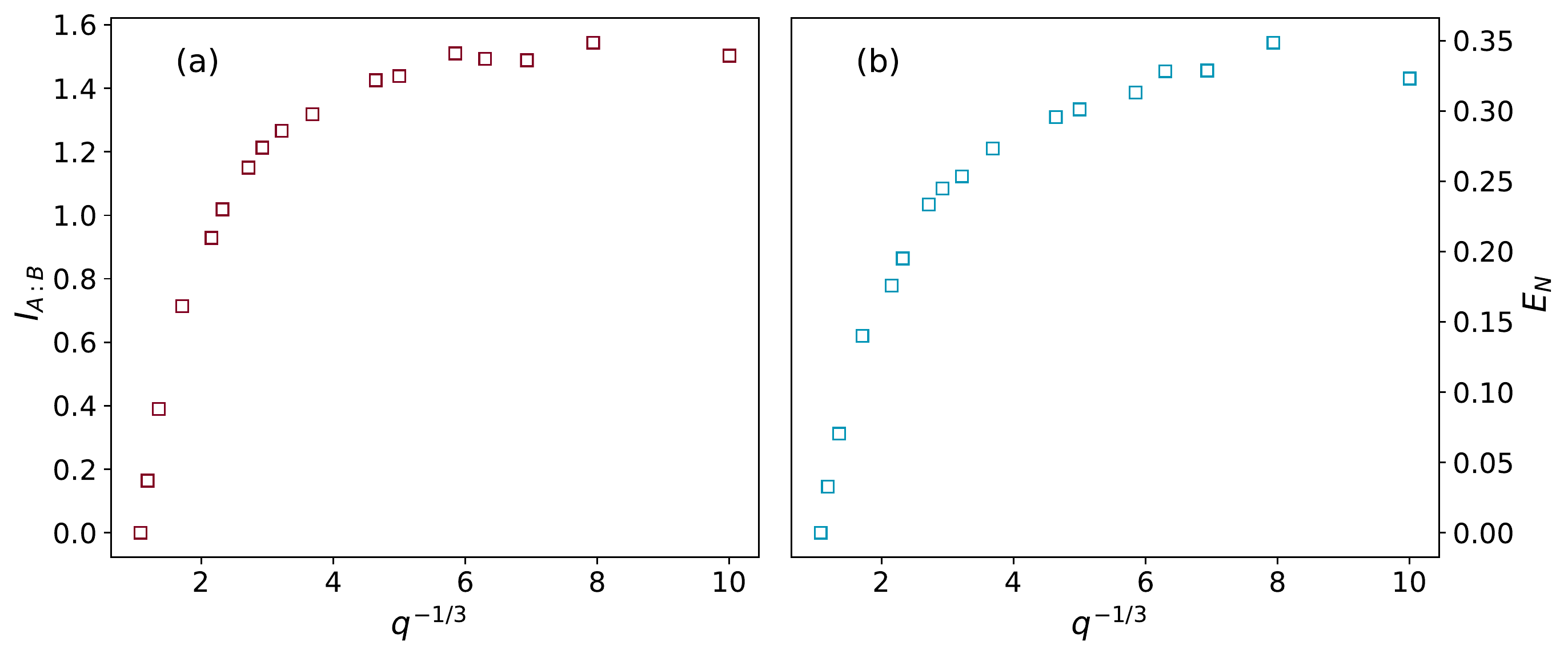}
	\caption{(a), The mutual information $I_{A:B}$ and (b), logarithmic entanglement negativity $E_{N}$ after extrapolating the system size to infinity (OBC). The measurement rate $p$ is zero.}
	\label{fig:pm0.0OBC}
	\end{figure}

	\subsection{D. Comparison of different fitting functions}
	We have also tried different fitting functions for the entanglement scaling against $q$, such as the log scaling $E_{N}(q) = a \log q + b$ (see Fig. \ref{fig:log}). Although the entanglement negativity seems to be linear with $\log q$ as shown in Fig. \ref{fig:log}(a) when $q < 0.004$, the $q^{-1/3}$ fit works better over a much larger $q$ regime. Moreover, the $q^{-1/3}$ power-law scaling can be understood as the KPZ fluctuations with emergent effective length scale $L_{\text{eff}} \sim q^{-1}$ as discussed below.

	\begin{figure}[H]\centering
	\includegraphics[width=0.5\textwidth]{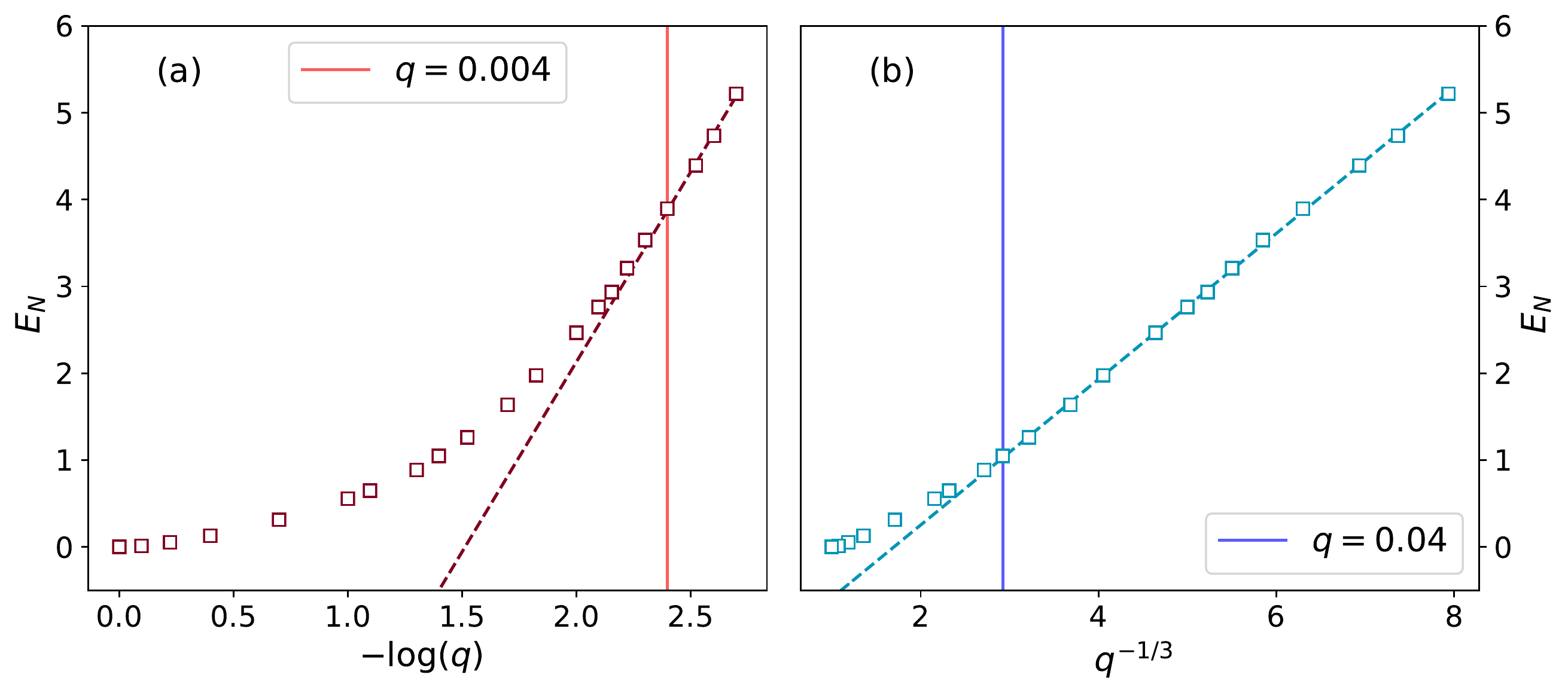}
	\caption{(a), The logarithmic entanglement negativity $E_{N}$ vs $\log(q)$ and (b), $E_{N}$ vs $q^{-1/3}$ (PBC). The measurement rate is $p=0.1$. The $q^{-1/3}$ scaling fits better over a much larger region compared to the log scaling hypothesis.}
	\label{fig:log}
	\end{figure}

	\subsection{E. Pinned phase via increasing the reset occurring rate $q$}
	In this section, we also consider the strategy by increasing the reset rate $q$ with fixed $t_{\text{noise}}$ on the top boundary to drive the system to enter the ``area law'' entanglement phase for the model with reset quantum channels on the last $t_{\text{noise}}$ layers (see Fig. \textcolor{red}{1}(b)). This model shares some similarities with the model considered in Ref. \cite{Pining_LYD}. The numerical results are shown in Fig. \ref{fig:tnoise4}. Here, we set $p=0.1$ and $t_{\text{noise}}=4$.

	When $q$ is small, the quantum noise is not strong enough to suppress the $O(L^{2/3})$ vertical fluctuations and the system exhibits the power law scaling entanglement ($L^{1/3}$); when $q$ is large, the system enters the pinned phase, i.e., ``area law'' entanglement phase \cite{Pining, Pining_LYD} (see the insets of the Fig. \ref{fig:tnoise4}(a)(b)). The logarithmic entanglement negativity and mutual information satisfy the universal scaling function proposed in \cite{QI_MIPT_PRX19}:
	\bea
	g(q,L)-g(q_c, L) = F((q-q_{c})L^{1/\nu}),
	\eea
	where $q_{c}$ is the critical rate, $\nu$ is the critical exponent which is related to the correlation length, and $F$ is an unknown function. The data collapse is shown in Fig. \ref{fig:tnoise4}. The critical rate is $q_{c}=0.035$ and the critical exponent is $\nu=0.94$, consistent with Ref. \cite{Noise_PhysRevLett22}. In the noise-driven ``area law'' phase ($q>q_{c}$), there is also a $q^{-1/3}$ scaling after extrapolating the system size to the thermodynamic limit as shown in Fig. \ref{fig:tnoise4}(c)(d).

	\begin{figure}[H]
	\centering
	\begin{minipage}[r]{0.5\textwidth}
	\centering
	\includegraphics[height=4.5cm,width=9cm]{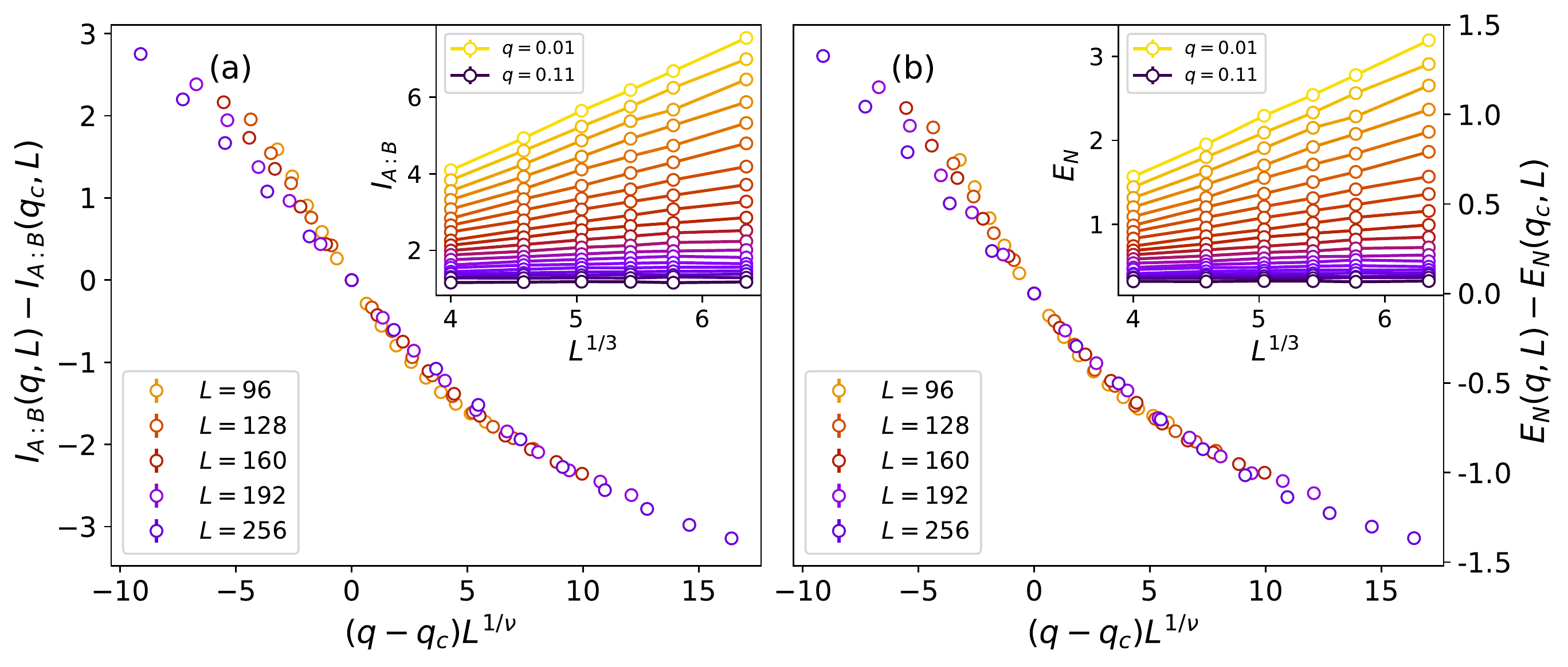}
	\end{minipage}%
	\begin{minipage}[c]{0.5\textwidth}
	\centering
	\includegraphics[height=4.5cm,width=9.5cm]{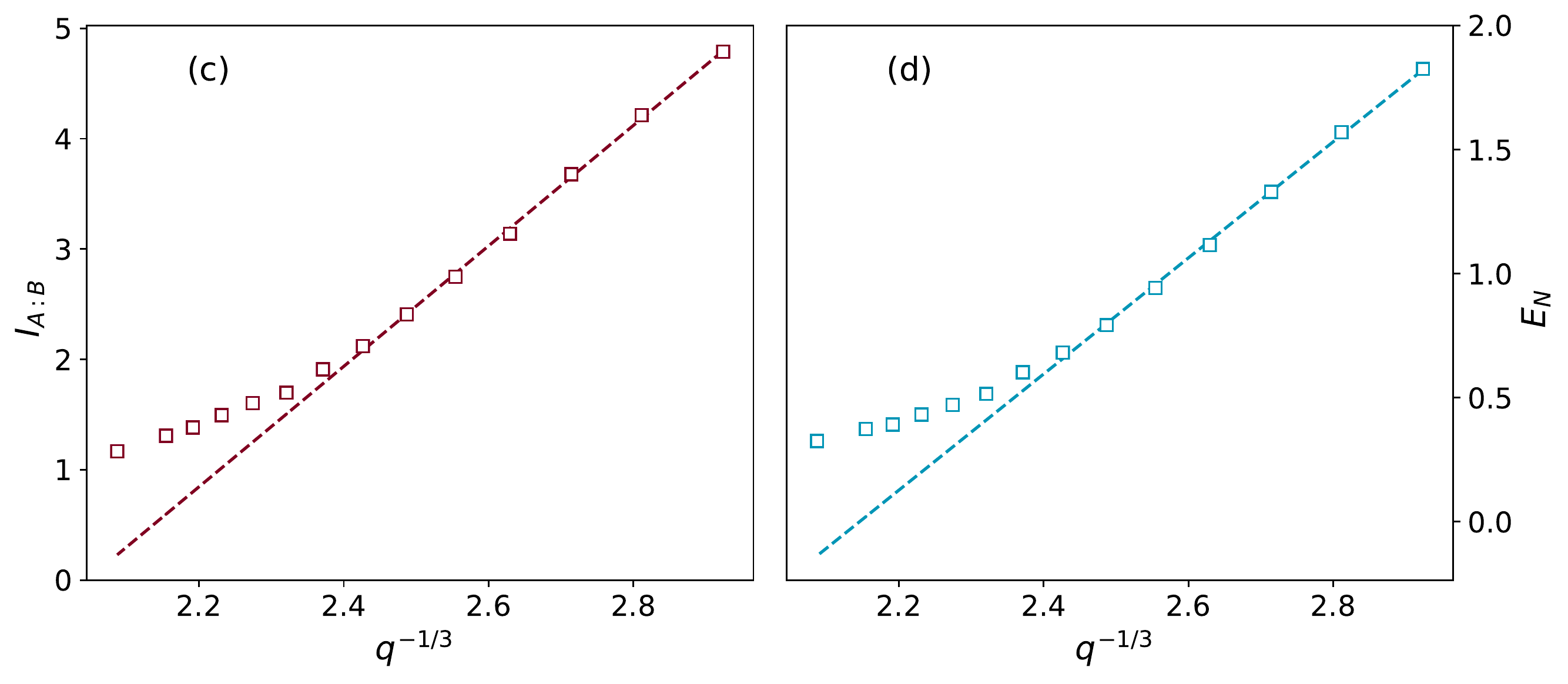}
	\end{minipage}
	\caption{Data collapse for (a), mutual information and (b), logarithmic entanglement negativity with $q \in [0.01,0.08]$ and fixed $t_{\rm{noise}}=4$. We report $q_{c}=0.035$ and $\nu=0.94$ from the finite size data collapse. With increasing the quantum noise rate $q$, the $L^{1/3}$ scaling disappears and the system enters the ``area law'' entanglement phase beyond the critical rate $q_{c}$ as shown in the insets of (a)(b). (c), The mutual information and (d), logarithmic entanglement negativity after extrapolating the system size to infinity obeys $q^{-1/3}$ scaling when $q$ moves above $q_{c}$ in the noise-driven ``area law'' entanglement phase.}
	\label{fig:tnoise4}
	\end{figure}

	\subsection{F. Clifford simulation}
	In this section, we review Clifford simulation utilized in the main text. For simplicity, we consider the case with $d=2$ qubits and a chain with $L$ qubits. In general, the total Hilbert space is:
	\bea
	H = H_{2}^{\otimes L},
	\eea
	which grows exponentially with the system size $L$ and is hard for classical simulation. However, for a special class of quantum states, we can utilize a set of Pauli strings ($P^{L}$) squaring to one to uniquely identify the quantum state. The set of Pauli strings forms an Abelian group known as the stabilizer group $\mathcal{G}$. The generators of the stabilizer group $\{g_{1}, g_{2}, ..., g_{\vert \mathcal{G} \vert}\}$ satisfy: 1, each $g_{i}$ is a product of Pauli operators; 2, $g^{\dagger}_{i}=g_{j}$ and $[g_{i}, g_{j}]=0$; 3, the set of generators are independent, i.e., $g_{i} \neq \prod_{j \neq i} g_{j}^{s_{j}}$, $s_{j}=0,1$. There are $2^{\vert \mathcal{G} \vert}$ elements, i.e., stabilizers, in the stabilizer group. The common eigenstate with $+1$ eigenvalue of these stabilizers is the stabilizer state. The purity of the stabilizer state is $\text{Tr}(\rho^2) = 2^{\vert \mathcal{G} \vert-L}$: the stabilizer state is a pure state for $\vert \mathcal{G} \vert = L$ and is a mixed state for $\vert \mathcal{G} \vert<L$. To understand the stabilizer description for quantum state, we consider a simple example: the product state $\vert 0 \rangle^{L}$ is the stabilizer state of the stabilizer group with generators $g_{i}=I^{\otimes L-1} \otimes Z_{i}$ for $i=1,..., L$, where $Z_{i}$ is the Pauli-Z matrix on $i$th qubit.

	The density matrix of the stabilizer state is:
	\bea
	\rho = \frac{2^{\vert \mathcal{G} \vert}}{2^{L}} \prod_{i=1}^{\vert \mathcal{G} \vert}(\frac{1+g_{i}}{2}) = \frac{1}{2^{L}} \sum_{g \in \mathcal{G}} g.
	\eea

	The stabilizer state is represented by $O(\vert \mathcal{G} \vert L)$ classical bits for the generators of $\mathcal{G}$ \cite{stabilizer_1998, stabilizer_PhysRevA} and thus much fewer computation resources are required to simulate evolution of such state. So long as $\rho$ remains a stabilizer state under the time evolution, the dynamics of $\rho$ can be efficiently simulated classically by keeping track of the evolution of the stabilizers $\mathcal{G}$. The set of unitary gates that map Pauli strings to Pauli strings and thus map stabilizer states to stabilizer states forms Clifford group:
	\bea
	\text{CI}(L) = \{ U \in \text{U}(L): UgU^{\dagger} \in P^{L} \ \text{for} \ g \in P^{L}\}.
	\eea
	Under the time evolution by a Clifford gate $U\in \text{CI}(L)$, the stabilizers $g$ evolve to $UgU^{\dagger}$. It is well known that the Clifford group can be universally generated by Hadamard, CNOT, and phase gates \cite{nielsen2002, stabilizer_PhysRevA}.

	% Measurement of Pauli string observables ($h$), including 
  The single qubit projective measurements ($h$) in the computational basis, considered in this work, also maps stabilizer states to stabilizer states. If $\rho$ is a mixed state, it is possible for $h$ to commute with all stabilizers in $\mathcal{G}$ without being an element of $\mathcal{G}$ itself. In this case, $\pm h$ is simply added as a new generator. More details can be found in Refs. \cite{stabilizer_PhysRevA, Noise_PhysRevLett22}.

	The reset quantum channel on qubit $l$, $R_{l}[\rho] = \sum_{a}E_{l}^{a} \rho E_{l}^{a\dagger}$, considered in the main text also preserves stabilizer states. The reset $R_{l}$ can be implemented by operating a swap operator between the qubit $l$ and an ancilla qubit initialized to the state $\vert 0 \rangle$.
	The new generators after adding the ancilla qubit are $\{g^{\prime}_{1}, g^{\prime}_{2}, ..., g^{\prime}_{\vert \mathcal{G} \vert}, g^{\prime}_{\vert \mathcal{G} \vert+1}\}$, where $g^{\prime}_{i} = g_{i} \otimes I_{aq} $ for $i=1,..., \vert \mathcal{G} \vert$ and $g^{\prime}_{\vert \mathcal{G} \vert+1} = I^{\otimes L} \otimes Z_{aq}$. The swap operator swaps the Pauli operators on qubit $l$ and the ancilla qubit, and maps $\{g^{\prime}_{1}, g^{\prime}_{2}, ..., g^{\prime}_{\vert \mathcal{G} \vert}, g^{\prime}_{\vert \mathcal{G} \vert+1}\}$ to  $\{g^{\prime \prime}_{1}, g^{\prime \prime}_{2}, ..., g^{\prime \prime}_{\vert \mathcal{G} \vert}, g^{\prime \prime}_{\vert \mathcal{G} \vert + 1}\} (\mathcal{G}_{aq})$. The density matrix after reset can be obtained by tracing out the ancilla qubit:
	\bea
	\rho^{\prime} = R_{l}[\rho] = \text{Tr}_{aq}(\rho_{aq}),
	\eea
	where $\rho_{aq} = \frac{2^{\vert \mathcal{G} \vert+1}}{2^{L+1}} \prod_{i=1}^{\vert \mathcal{G} \vert+1}(\frac{1+g^{\prime \prime}_{i}}{2}) = \frac{1}{2^{L+1}} \sum_{g_{aq} \in \mathcal{G}_{aq}} g_{aq}$. $\mathcal{G}^{\prime} = \{ g \vert g \otimes I_{aq} \in \mathcal{G}_{aq} \}$ (the Pauli matrices $X,Y,Z$ are traceless) is a subgroup of $\mathcal{G}_{aq}$ and is the stabilizer group corresponding to $\rho^{\prime}$. The purity is $\text{Tr}(\rho^{\prime 2}) = 2^{\vert \mathcal{G}^{\prime} \vert -L}$, where $\vert \mathcal{G}^{\prime} \vert$ is the number of generators of $\mathcal{G}^{\prime}$. Different from the dephasing channel considered in Ref. \cite{Noise_PhysRevLett22}, reset quantum channel can also increase the purity. For example,
	\bea
	R_{1}[R_{2}[...R_{L}[\rho]]] = \vert 0 \rangle \langle 0 \vert ^{ \otimes L}
	\eea
	for any $\rho$ and the purity is $1$ now.

	The logarithmic entanglement negativity $E_{N}$ and the mutual information $I_{A:B}$ can also be easily obtained from the stabilizers. For subsystem $A$ with $L_{A}$ qubits and its complementary subsystem $B$ with $L_{B}=L-L_{A}$ qubits,
	\bea
	\rho_{A} &=& \text{Tr}_{B}(\rho) = \frac{1}{2^{L_{A}}} \sum_{g_{A} \in \mathcal{G}_{A}} g_{A} , \\
	\rho_{B} &=& \text{Tr}_{A}(\rho) = \frac{1}{2^{L_{B}}} \sum_{g_{B} \in \mathcal{G}_{B}} g_{B}
	\eea
	where $\mathcal{G}_{A} = \{ g_{A} \vert g_{A}\otimes I_{B} \in \mathcal{G} \}$ and $\mathcal{G}_{B} = \{ g_{B} \vert g_{B}\otimes I_{A} \in \mathcal{G} \}$. The entanglement entropy $S_{A} = \vert \mathcal{G}_{A} \vert - L_{A}$, $S_{B} = \vert \mathcal{G}_{B} \vert - L_{B}$, and $S_{AB} = \vert \mathcal{G} \vert - L$. And the mutual information:
	\bea
	I_{A:B} = S_{A} + S_{B} - S_{AB} = \vert \mathcal{G}_{A} \vert + \vert \mathcal{G}_{B} \vert - \vert \mathcal{G} \vert.
	\eea
	To obtain the logarithmic negativity $E_{A:B}$, we can define a $\vert \mathcal{G} \vert \times \vert \mathcal{G} \vert$ symmetric matrix $J$:
	\bea
	J_{ij} = \left\{
	\begin{aligned}
	&1&  \ \{g_{A}^{i}, g_{A}^{j}\} = 0,  \\
	&0&  \ \text{otherwise},
	\end{aligned}
	\right.
	\eea
	where $g^{i}=g^{i}_{A} \otimes g^{i}_{B}$. $E_{N} = \frac{1}{2} \text{rank}(J)$ over field $F_{2}$ \cite{Negativity_PRXQuantum21, MI_EN_Dai}. When we choose the following set of generators of $\mathcal{G}$:
	\bea
	\{ g_{A}^{i} \otimes I_{B} \}_{i=1}^{\vert \mathcal{G}_{A} \vert} \cup \{ I_{A} \otimes g^{j}_{B} \}_{j=1}^{\vert \mathcal{G}_{B} \vert} \cup \{ g_{A}^{k} \otimes g^{k}_{B} \}_{k=1}^{\vert \mathcal{G} \vert - \vert \mathcal{G}_{A} \vert - \vert \mathcal{G}_{B} \vert},
	\eea
	with $J = (0)^{\oplus \vert \mathcal{G}_{A} \vert} \oplus (0)^{\oplus \vert \mathcal{G}_{B} \vert} \oplus J^{\prime}$. It is obvious that $E_{N} \leq \frac{1}{2} I_{A:B}$ because that $\text{rank}(J)=\text{rank}(J^{\prime}) \leq \vert \mathcal{G} \vert - \vert \mathcal{G}_{A} \vert - \vert \mathcal{G}_{B} \vert$. It should be noted that this bound is for the stabilizer state and can not be generalized to arbitrary quantum states \cite{MI_EN_Dai}.

	\subsection{G. Effective statistical model}
	\subsubsection{Replica entropy}
	In this section, we will introduce how to obtain the entanglement entropy and mutual information via replica entropy (more details on the mapping framework can be found in Ref. \cite{Noise_PhysRevLett22}).

	With fixed sets of measurement locations $X$, reset locations $Y$, the unitary realization $U$, and measurement history trajectories $m$, the unnormalized density matrix at time step $t$ (see Fig. \textcolor{red}{1}(a)) is given by
	\bea
	\rho_{m,X,Y} = P_{t} R_{t} \left[  U_{t}...P_{1} R_{1}\left[ U_{1} \rho_{0} U_{1}^{\dagger} \right] P^{\dagger}_{1}...U^{\dagger}_{t}\right]P^{\dagger}_{t},
	\eea
	where $\rho_{0} = \vert 0 \rangle \langle 0 \vert^{\otimes L}$, $U_{i}$ is the product of two-qudit unitary gates within the $i$th layer, $R_{i}$ is the reset quantum channel on $i$th layer, and $P_{i}$ denotes the projective measurements on $i$th layer.
 % projects onto the outcomes of measurements performed after the $i$th layer.

	The $n$th R$\rm{\acute{e}}$nyi entropy $S_{\alpha}^{(n)}$ ($\alpha=A,B,AB$) averaged over unitary realizations $U$ is given by
	\bea
	\overline{S_{\alpha}^{(n)}(X,Y)} &=& \mathbb{E}_{U} \sum_{m}p_{m,X,Y} \frac{1}{1-n} \rm{log} \left\{ \frac{\rm{tr} \rho^{n}_{\alpha,m,X,Y}}{(\rm{tr}\rho_{m,X,Y})^{n}} \right\} \\ \nonumber
	&=& \mathbb{E}_{U} \sum_{m}p_{m,X,Y} \frac{1}{1-n} \rm{log} \left\{ \frac{Z_{S_{\alpha}}^{(n)}}{Z^{(n)}}\right\},
	\eea
	where $\rho_{\alpha, m, X,Y} = \text{tr}_{\bar{\alpha}} \rho_{m,X,Y}$ ($\bar{\alpha}$ is the complementary subsystem of subsystem $\alpha$), $p_{m,X,Y}=\text{tr} \rho_{m,X,Y}$ is the probability for this specific configuration. 
 % achieving the measurement outcomes $m$ conditioned on the locations of measurements $X$, the locations of resets $Y$ and the unitary realization $U$.
	We can obtain the entanglement entropy $S_{\alpha}=\lim_{n \rightarrow 1} S_{\alpha}^{(n)}$ and mutual information $I_{A:B}=\lim_{n \rightarrow 1} S_{A}^{(n)} + S_{B}^{(n)} - S_{AB}^{(n)}$. The average outside the $\log$ function is difficult and we employ the replica trick \cite{replica1, replica2} to perform the average over unitary realizations inside the log,
	\bea
	\mathbb{E}_{U} \sum_{m}p_{m,X,Y} \rm{log} Z_{S_{\alpha}}^{(n)} = \rm{lim}_{k \rightarrow 0} \frac{1}{k} \rm{log} \left\{ \mathbb{E}_{U} \sum_{m}p_{m,X,Y} (Z_{S_{\alpha}}^{(n)})^{k} \right\} = \rm{lim}_{k \rightarrow 0} \frac{1}{k} \rm{log} Z_{S_{\alpha}}^{(n,k)},
	\eea
	where
	\bea
	Z_{S_{\alpha}}^{(n,k)} = \mathbb{E}_{U} \sum_{m}p_{m,X,Y} (Z_{S_{\alpha}}^{(n)})^{k} = \mathbb{E}_{U} \sum_{m} \rm{Tr}[(\Sigma_{S_{\alpha}}^{\otimes k} \otimes \mathbb{I}) \rho_{m,X,Y}^{\otimes nk+1}] = \rm{Tr} \left\{  (\Sigma_{S_{\alpha}}^{\otimes k }\otimes \mathbb{I})[\mathbb{E}_{U} \sum_{m} \rho_{m,X,Y}^{\otimes r}]\right\}.
	\eea
	And
	\bea
	Z^{(n,k)} = \rm{Tr} \left\{  \mathbb{I}^{\otimes r}[\mathbb{E}_{U} \sum_{m} \rho_{m,X,Y}^{\otimes r}]\right\}.
	\eea
	The difference between different partition functions is the different permutation freedom $\Sigma$ on the top boundary as shown in Fig. \textcolor{red}{3}. The R$\rm{\acute{e}}$nyi entropy $S_{\alpha}^{(n)}$ equals to the replica entropy $S_{\alpha}^{(n,k)}$ in the $k \rightarrow 0$ limit, where
	\bea
	S_{\alpha}^{(n,k)}(X,Y) = \frac{1}{k(1-n)} \rm{log} \left\{ \frac{Z_{S_{\alpha}}^{(n,k)} (X,Y)}{Z^{(n,k)}(X,Y)} \right\}.
	\eea
	The replica entropy is thus represented as the free energy difference of the effective statistical models with different boundary conditions,
	\bea
	S_{\alpha}^{(n,k)}(X,Y) = \frac{1}{k(n-1)} [F_{S_{\alpha}}^{(n,k)}(X,Y) - F^{(n,k)}(X,Y)].
	\eea
	
	\subsubsection{Free energy}
	In this section, we will introduce the calculation of the free energy of the effective statistical model. In the large $d \rightarrow \infty$ limit, the calculation is simplified and the free energy is determined by the most probable classical spin configuration.

 	The average over each Haar unitary $U_{t,ij}$ yields \cite{QI3_Nahum, PhysRevB.99.174205, MIPT_Clifford_PRB20_Bao, MIPT_Clifford_PRB21_Jian, PhysRevX.12.041002, collins2003moments, collinsIntegrationRespectHaar2006, PhysRevX.8.021014, Noise_PhysRevLett22}
 	\bea
 	\rm{\mathbb{E}}_{U}(U_{t,ij} \otimes U_{t,ij}^{*})^{\otimes r} = \sum_{\sigma, \tau \in S_{r}} \rm{Wg}_{\rm{d^{2}}}^{(r)}(\sigma \tau^{-1}) \vert \tau \tau \rangle \langle \sigma \sigma \vert_{ij},
 	\eea
 	where $\rm{Wg}_{\rm{d^{2}}}^{(r)}(\sigma)$ is the Weingarten function and has asymptotic expansion for large $d$ as follows \cite{ PhysRevB.99.174205, collinsIntegrationRespectHaar2006}:
 	\bea
 	\rm{Wg}_{\rm{d^{2}}}^{(r)}(\sigma) = \frac{1}{d^{2r}} [\frac{\rm{Moeb}(\sigma)}{d^{2\vert \sigma \vert}} + O(d^{-2 \vert \sigma \vert -4})],
 	\eea
 	where $\vert \sigma \vert$ is the number of transpositions required to build $\sigma$ from the identity permutation $\rm{\mathbb{I}}$ and Moeb($\sigma$) is the Moebius number of $\sigma$ \cite{collinsIntegrationRespectHaar2006}. The permutation-valued spins $\sigma$ and $\tau$ form the classical degrees of freedom for the analytical statistic model.

 	Between two diagonally neighboring permutation spins, in the absence of measurement, the tensor contraction is given by
 	\bea
 	w_d(\sigma, \tau) = \langle \sigma \vert \tau \rangle = d^{r-\vert \sigma^{-1} \tau \vert}.
 	\eea
 	In the presence of a measurement, the weight is given by
 	\bea
 	w^{m}_{d}(\sigma, \tau) = \langle \sigma \vert (P^{a} \otimes P^{a})^{\otimes r}  \vert \tau \rangle = 1,
 	\eea
 	which is independent on $\sigma$ and $\tau$.
 	If a reset is present, the weight is
 	\bea
 	w^{R}_{d}(\sigma, \tau) = \langle \sigma \vert R^{\otimes r} \vert \tau \rangle = d^{r - \vert \tau \vert},
 	\eea
 	which is independent on permutation spin $\sigma$ and is maximized as $\tau = \mathbb{I}$.
 	If a measurement and a reset are present,
 	\bea
 	w^{R,m}_{d}(\sigma, \tau) = \langle \sigma \vert (P^{a} \otimes P^{a})^{\otimes r} R^{\otimes r} \vert \tau \rangle  =\delta_{a,0} d^{r - \vert \tau \vert}.
 	\eea	
 	When $a \neq 0$, this trajectory can be removed from the weight summation; when $a = 0$, it is the same as the case with only resets present.

 	Further simplification arises when the $\tau$ spins are traced out.  We can obtain effective three-body interaction of downward-facing triangles in the absence of measurements and reset quantum channels, which is given by
 	\bea
 	W^{0}(\sigma_1, \sigma_2; \sigma_3) = \sum_{\tau \in S_{r}} \rm{Wg}_{d^2}^{(r)}(\sigma_3 \tau^{-1}) d^{2r - \vert \sigma_{1}^{-1}\tau \vert - \vert \sigma_{2}^{-1}\tau \vert},
 	\eea
 	with a unitary constrain,
  	\bea
  	\label{constrain}
 	W^{0}(\sigma, \sigma; \sigma_{3}) = \sum_{\tau \in S_{r}} \rm{Wg}_{d^2}^{(r)}(\sigma_3 \tau^{-1})(d^{2})^{r-\vert \sigma^{-1} \tau \vert} = \delta_{\sigma, \sigma_{3}},
 	\eea
 	which indicates that there is no horizontal domain wall in such triangles in the absence of measurements and resets. In large $d$ limit,
 	\bea
 	\label{W0}
 	W^{0}(\sigma^{\prime}, \sigma; \sigma) = W^{0}(\sigma, \sigma^{\prime}; \sigma) =&& \sum_{\tau \in S_{r}} \rm{Wg}_{d^2}^{(r)}(\sigma \tau^{-1}) d^{2r - \vert \sigma^{-1}\tau \vert - \vert \sigma^{\prime -1}\tau \vert} \\ \nonumber
 	\approx && \sum_{\tau \in S_{r}} \rm{Moeb}(\sigma \tau^{-1}) d^{-2 \vert \sigma \tau^{-1} \vert - \vert \sigma^{-1} \tau \vert -  \vert \sigma^{\prime -1} \tau \vert} \\ \nonumber
 	\overset{d \rightarrow \infty}{=} && d^{-\vert \sigma^{-1} \sigma^{\prime} \vert} \ (\tau = \sigma).
 	\eea

	In the presence of a measurement, assuming it is between $\sigma_1$ and $\sigma_3$,
 	\bea
 	W^{m}(\sigma_2; \sigma_3) &=& \sum_{\tau \in S_{r}} \rm{Wg}_{d^2}^{(r)}(\sigma_3 \tau^{-1}) d^{r - \vert \sigma_{2}^{-1}\tau \vert} \\ \nonumber
 	& \approx & \sum_{\tau \in S_{r}} \frac{1}{d^{2r}} \frac{\rm{Moeb}(\sigma_{3} \tau^{-1})}{d^{2\vert \sigma_{3} \tau^{-1} \vert}} d^{r - \vert \sigma_{2}^{-1}\tau \vert} \\ \nonumber
 	& \overset{d \rightarrow \infty}{=} & \frac{1}{d^{r}} d^{-\vert \sigma_{2}^{-1} \sigma_{3} \vert} \ \rm{with} \ \tau=\sigma_{3}.
 	\eea
 	We can add an unimportant factor: $W^{m}(\sigma_2; \sigma_3) \overset{d \rightarrow \infty}{=} d^{-\vert \sigma_{2}^{-1} \sigma_{3} \vert}$. We can see that the $\sigma_{1}$ is decoupled from the other two permutation spins $\sigma_{2}$ and $\sigma_{3}$. In this case, if there is a domain wall between $\sigma_{1}$ and other two permutation spins ($\sigma_{1} = \sigma^{\prime}$ and $\sigma_{2} = \sigma_{3} =  \sigma$), $W^{m}(\sigma^{\prime}, \sigma; \sigma) \overset{d \rightarrow \infty}{=} 1$ which is smaller than $	W^{0}(\sigma^{\prime}, \sigma; \sigma)$ (see Eq. \ref{W0}).

 	The weights without resets discussed above only depend on $\vert \sigma^{-1} \sigma^{\prime} \vert$ and the weights are invariant under transformations of the form
 	\bea
 	\sigma \mapsto \xi_{1} \sigma \xi_{2}^{-1}, \ \  \sigma^{\prime} \mapsto \xi_{1} \sigma^{\prime} \xi_{2}^{-1},
 	\eea
 	which means swapping all ket and bra indices independently. And the weights are invariant under inversion,
 	\bea
 	\sigma \mapsto \sigma^{-1}, \ \ \sigma^{\prime} \mapsto \sigma^{\prime -1}.
 	\eea
 	Therefore, the symmetry group is $(S_{r} \times S_{r}) \rtimes \mathbb{Z}_{2}$.

 	The partition function is equal to the product of the weights ($d^{-\vert \sigma^{-1} \sigma^{\prime}} \vert$) and the contribution to the free energy is:
 	\bea
 	\beta E(\sigma, \sigma^{\prime}) = \vert \sigma^{-1} \sigma^{\prime} \vert \log d.
 	\eea
 	Then the total free energy is equal to the product of the domain wall length and $\vert \sigma^{-1} \sigma^{\prime} \vert$, where $\sigma$ and $\sigma^{\prime}$ are the permutations spins separated by the domain wall. The domain wall length is required to be the shortest in the
 	 most probable spin configuration with the lowest free energy. The shortest domain wall is unique with zero measurements. If a domain wall crosses the bond with a measurement, the free energy contribution is zero thus the domain wall goes through as many as possible measurements to reduce the free energy and slightly fluctuates away from the unique trajectory. In a coarse-grained picture with some approximations \cite{Noise_PhysRevLett22}, the measurements with random locations can be regarded as the Gaussian attractive potential. We can view the domain wall as the directed polymer and the free energy of a directed polymer in a random Gaussian potential satisfies the KPZ equation \cite{KPZ1, KPZ2, KPZ3,Noise_PhysRevLett22}.

 	In the presence of a reset quantum channel, assuming it is between $\sigma_1$ and $\sigma_3$,
 	\bea
 	W^{R}(\sigma_2; \sigma_3) &=& \sum_{\tau \in S_{r}} \rm{Wg}_{d^2}^{(r)}(\sigma_3 \tau^{-1}) d^{2r - \vert \tau \vert - \vert \sigma_{2}^{-1}\tau \vert} \\ \nonumber
 	& \approx & \sum_{\tau \in S_{r}} \frac{1}{d^{2r}} \frac{\rm{Moeb}(\sigma_{3} \tau^{-1})}{d^{2\vert \sigma_{3} \tau^{-1} \vert}} d^{2r - \vert \tau \vert - \vert \sigma_{2}^{-1}\tau \vert} \\ \nonumber
 	& = & \sum_{\tau \in S_{r}} \rm{Moeb}(\sigma_{3} \tau^{-1}) d^{-2\vert \sigma_{3} \tau^{-1} \vert - \vert \tau \vert - \vert \sigma_{2}^{-1}\tau \vert}.
 	\eea
 	The $\sigma_{1}$ is decoupled from the other two permutation spins $\sigma_{2}$ and $\sigma_{3}$ similar to the case with a measurement. However,
 	\bea
 	W^{R}(\sigma^{\prime}, \sigma; \sigma) \approx \sum_{\tau \in S_{r}} \rm{Moeb}(\sigma \tau^{-1}) d^{-3\vert \sigma \tau^{-1} \vert - \vert \tau \vert } \overset{d \rightarrow \infty}{=} d^{-\vert \sigma \vert} \ (\tau = \sigma),
 	\eea
 	with $\sigma_{1} = \sigma^{\prime}$ and $\sigma_{2} = \sigma_{3} =  \sigma$. The weight is minimized as the permutation spin $\sigma = \mathbb{I}$, otherwise, there is additional energy cost. As shown in Fig. \textcolor{red}{3}, there are two different permutation spins $\mathbb{C}$ and $\mathbb{I}$. The additional free energy contribution is proportional to $N_{R} \vert \mathbb{C} \vert$, where $N_{R}$ is the number of bulk resets in the region of the domain $\mathbb{C}$. Equivalently, the bulk resets can be regarded as the top boundary attractive potential, which tends to push the domain wall to the top boundary, i.e., suppresses the large-scale vertical fluctuations of the directed polymers. In the presence of quantum channels, the unitary constrain (see Eq. \ref{constrain}) fails and it is possible that the domain wall passes through the triangle horizontally.

 	For the permutation spin $\tau$ adjacent to the reset quantum channel at the top boundary or the side boundary, $\tau$ is only coupled to another permutation spin $\sigma$ via a vertical bond, instead of the case adjacent to the bulk reset where $\tau$ is coupled to two permutation spins as discussed above, with the weight
 	\bea
 	W^{R,b}(\tau, \sigma) & = & \sum_{\tau \in S_{r}} \rm{Wg}_{d^2}^{(r)}(\sigma \tau^{-1}) d^{r-\vert \tau \vert} \\ \nonumber
 	& \approx & \sum_{\tau \in S_{r}} \frac{1}{d^{2r}} \frac{\rm{Moeb}(\sigma \tau^{-1})}{d^{2\vert \sigma \tau^{-1} \vert}} d^{r-\vert \tau \vert},
 	\eea
 	which is maximized with $\sigma = \tau = \mathbb{I}$. In the large $d$ limit, the spin $\sigma$ is pinned to the identity spin $\mathbb{I}$. The resets near the top spatial boundary can induce an effective length scale $L_{\text{eff}} \sim  q^{-1}$ as indicated in Fig. \textcolor{red}{3}%\ref{fig:MI}
 	. As discussed in the next section, the KPZ fluctuations with the emergent effective length scale $L_{\text{eff}} \sim  q^{-1}$ cause the novel power law scaling entanglement in terms of the quantum channel occurring probability $q$.

	It is worth noting that all decoherence quantum channels, not just reset channels, couple ket and bra within individual copies so that the weights only remain invariant under permutations of the $r$ copies of the qudit and under Hermitian conjugation. Thus the symmetry group $(S_{r} \times S_{r}) \rtimes \mathbb{Z}_{2}$ is broken into $S_{r} \rtimes \mathbb{Z}_{2}$.

 	\subsubsection{$q^{-1/3}$ scaling for mutual information}
 	The mutual information is the free energies difference of the effective statistical models with different boundary conditions as indicated in Fig. \textcolor{red}{3}%\ref{fig:MI}
 	. And the free energy is proportional to the directed polymer length in a random attractive potential with an emergent effective length scale $L_{\text{eff}} \sim q^{-1}$ as shown in Fig. \textcolor{red}{3} %\ref{fig:MI}.
 	\bea
 	\label{eq:MI}
 	I_{A:B} &=&  \underset{\underset{k \rightarrow 0}{n \rightarrow 1}}{\lim} \frac{1}{k(n-1)} (F^{(n,k)}_{S_{A}} + F^{(n,k)}_{S_{B}} - F^{(n,k)}_{S_{AB}}) \\ \nonumber
 	&=& \underset{\underset{k \rightarrow 0}{n \rightarrow 1}}{\lim} \frac{1}{k(n-1)} \vert \mathbb{CI} \vert ( l_{S_{A}} +  l_{S_{B}} - l_{S_{AB}}) \\ \nonumber
 	&=& l_{S_{A}} + l_{S_{B}} - l_{S_{AB}} \\ \nonumber
 	&=& s_{0} (\frac{L_{\text{eff}}}{2}) + s_{1} (\frac{L_{\text{eff}}}{2})^{1/3} + s_{0} (\frac{L_{\text{eff}}}{2}) + s_{1} (\frac{L_{\text{eff}}}{2})^{1/3} - s_{0} L_{\text{eff}} - s_{1} L_{\text{eff}}^{1/3}, \\ \nonumber
 	&=& 2s^{\prime}_{1}(\frac{q^{-1}}{2})^{1/3} - s^{\prime}_{1} q^{-1/3} \\ \nonumber
 	&=& 2s^{\prime}_{1}(\frac{1}{2^{1/3}} -\frac{1}{2}) q^{-1/3}.
 	\eea
 	The contribution to the mutual information is from the directed polymer with two endpoints in regions $A$ and $B$ respectively as the middle directed polymer shown in Fig. \textcolor{red}{3}%\ref{fig:MI}
 	(c). Here, we assume the length scales of the directed polymer for $Z^{(n,k)}_{S_{A}}$ and $Z^{(n,k)}_{S_{B}}$, which has a nonzero contribution to the mutual information, are $l_{A}^{\prime} = l_{B}^{\prime} = \frac{L_{\text{eff}}}{2}$ for simplicity. For the case $l_{A}^{\prime} \neq l_{B}^{\prime}$, the contribution to the mutual information is also proportional to $q^{-1/3}$.

 	\subsubsection{$q^{-1/3}$ scaling for logarithmic entanglement negativity}

 	The calculation of the logarithmic entanglement negativity which can be obtained from the replica negativity is similar to that of the mutual information,
 	\bea
 	E_{N} &=&  \underset{\underset{k \rightarrow 0}{n \rightarrow 1}}{\lim} \frac{1}{k(2-n)}\log(\frac{Z_{E_{N}}^{(n,k)}}{Z_{E_{N0}}^{(n,k)}}) \\ \nonumber
 	&=&  \underset{\underset{k \rightarrow 0}{n \rightarrow 1}}{\lim} \frac{1}{k(n-2)} (F_{E_{N}}^{(n,k)} - F^{(n,k)}_{E_{N0}}).
 	\eea
 	As reported in Ref. \cite{Noise_PhysRevLett22},  	there can be an intermediate domain where the spin is $\mathbb{D}$ as shown in Fig. \ref{fig:EN}, which satisfy
 	\bea
    \vert \mathbb{C}^{-1} \mathbb{D} \vert &+& \vert \mathbb{D} \mathbb{I} \vert = \vert \mathbb{\bar{C}}^{-1} \mathbb{D} \vert + \vert \mathbb{D} \mathbb{I} \vert = \vert \mathbb{C} \mathbb{I} \vert, \\
    \vert \mathbb{C}^{-1} \mathbb{D} \vert &+& \vert \bar{\mathbb{C}}^{-1} \mathbb{D} \vert = \vert \mathbb{C}^{-1} \bar{\mathbb{C}} \vert.
 	\eea
 	Then
 	\bea
 	\vert \mathbb{C}^{-1} \mathbb{D} \vert = \vert \mathbb{\bar{C}}^{-1} \mathbb{D} \vert = k(\frac{n}{2}-1), \ \ \vert \mathbb{D} \mathbb{I} \vert = k \frac{n}{2}.
 	\eea
 	The existence of the intermediate domain $\mathbb{D}$ can further reduce the free energy. The boundary conditions of $Z^{(n,k)}_{E_{N}}$ and $Z^{(n,k)}_{E_{N0}}$ are shown in Fig. \ref{fig:EN}. The entanglement negativity is
 	\bea
 	\label{eq:EN}
 	E_{N} &=& \underset{\underset{k \rightarrow 0}{n \rightarrow 1}}{\lim}\frac{1}{k(n-2)} (\vert \mathbb{C}^{-1} \mathbb{D} \vert l_{\mathbb{CD}} + \vert \mathbb{\bar{C}}^{-1} \mathbb{D} \vert l_{\mathbb{\bar{C}D}} + \vert \mathbb{D} \mathbb{I} \vert l_{\mathbb{DI}} - \vert \mathbb{C}^{-1} \mathbb{I} \vert l_{\mathbb{CI}}) \\ \nonumber
 	&=& \underset{\underset{k \rightarrow 0}{n \rightarrow 1}}{\lim} \frac{1}{k(n-2)} (2k(\frac{n}{2} - 1) (s_{0} (\frac{L_{\text{eff}}}{2})  + s_{1} (\frac{L_{\text{eff}}}{2})^{1/3} ) + k \frac{n}{2}(s_{0} L_{\text{eff}} + s_{1} (L_{\text{eff}})^{1/3}) - k(n-1)(s_{0} L_{\text{eff}} + s_{1} (L_{\text{eff}})^{1/3}) ) \\ \nonumber
 	&=& s_{1}^{\prime} (\frac{1}{2^{1/3}} - \frac{1}{2}) q^{-1/3} \\ \nonumber
 	&=& \frac{1}{2} I_{A:B}
 	\eea

 	\begin{figure}[H]\centering
		\includegraphics[width=0.5\textwidth]{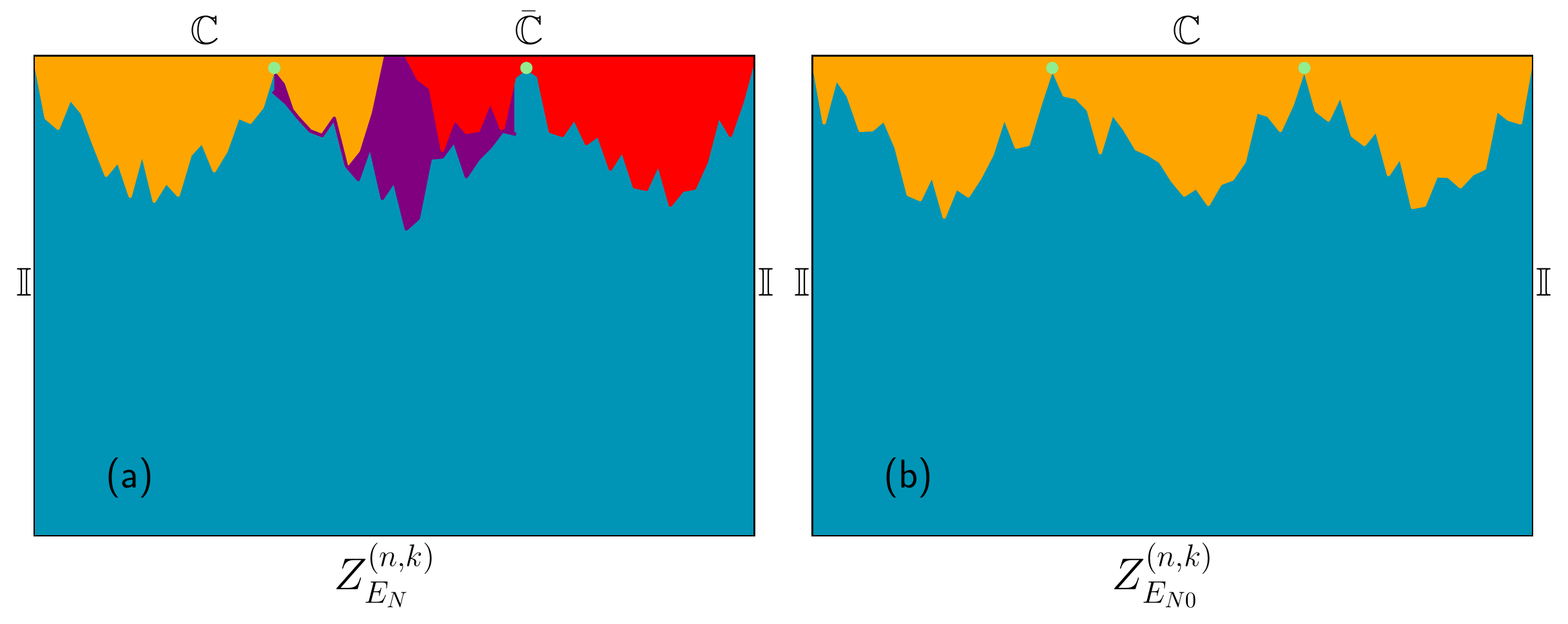}
		\caption{The domain configuration in the presence of resets (green dots): the domain configurations for (a), $Z_{E_{N}}^{(n,k)}$, (b), $Z_{E_{N0}}^{(n,k)}$. The purple region represents the domain in which the permutation spin is $\mathbb{D}$.}
		\label{fig:EN}
	\end{figure}

 	\subsubsection{Breakdown of KPZ theory with large $q$}
 	We have demonstrated that the $q^{-1/3}$ scaling entanglement can be understood as the KPZ fluctuations of the directed polymer with an emergent effective length scale $L_{\text{eff}} \sim q^{-1}$. The length of the directed polymer
 	\bea
 	l = s_{0} L_{\text{eff}} + s_{1} L_{\text{eff}}^{1/3},
 	\eea
 	utilized in Eqs. \ref{eq:MI} and \ref{eq:EN}, is only valid for large $L_{\text{eff}}$ \cite{KPZ_largeL} since KPZ theory is a field theory valid in the continuum limit. When the effective length scale is of the same order as the discrete lattice constant, the field theory description breaks down. As shown in Fig. \textcolor{red}{2}%\ref{fig:pm0.1PBC}
 	, when $q$ is close to $1$, i.e., $L_{\text{eff}}$ is small ($L_{\text{eff}}<20$), the entanglement value deviates from the prediction scaling $q^{-1/3}$  based on KPZ theory. Besides, we also observe that the breakdown of the relation $E_N = \frac{1}{2} I_{A:B}$ as shown in Fig.~\ref{fig:pm0.1OBC} and the breakdown of the plateau with varying $q$ for large $p$ as shown in Fig.~\ref{fig:largep} both happen near the same $q$ value indicating the breakdown of the field theory approximation.  Therefore, we draw the conclusion that these phenomena mentioned above when $q$ is large are not of central interest since they are not universal as indicated by the break-down of field theory description.

\end{widetext}

\end{document}